# Atomistic-continuum modeling of short laser pulse melting of Si targets

V.P. Lipp[1,2], B. Rethfeld[1], M.E. Garcia[2], D.S. Ivanov[1,2]

[1]*Department of Physics and OPTIMAS Research Center,*

*Technical University of Kaiserslautern, 67663 Kaiserslautern*

[2]*Theoretical Physics (FB10), University of Kassel, Heinrich-Plett-Strasse 40, 34132 Kassel, Germany*

## Abstract

We present an atomistic-continuum model to simulate ultrashort laser-induced melting processes in semiconductor solids on the example of silicon. The kinetics of transient non-equilibrium phase transition mechanisms is addressed with a Molecular Dynamics method at atomic level, whereas the laser light absorption, strong generated electron-phonon non-equilibrium, fast diffusion and heat conduction due to photo-excited free carriers are accounted for in the continuum. We give a detailed description of the model, which is then applied to study the mechanism of short laser pulse melting of free standing Si films. The effect of laser-induced pressure and temperature of the lattice on the melting kinetics is investigated. Two competing melting mechanisms, heterogeneous and homogeneous, were identified. Apart of classical heterogeneous melting mechanism, the nucleation of the liquid phase homogeneously inside the material significantly contributes to the melting process. The simulations showed, that due to the open diamond structure of the crystal, the laser-generated internal compressive stresses reduce the crystal stability against the homogeneous melting. Consequently, the latter can take a massive character within several picoseconds upon the laser heating. Due to negative volume of melting of modeled Si material, -7.5%, the material contracts upon the phase transition, relaxes the compressive stresses and the subsequent melting proceeds heterogeneously until the excess of thermal energy is consumed. The threshold fluence value, at which homogeneous nucleation of liquid starts contributing to the classical heterogeneous propagation of the solid-liquid interface, is found from the series of simulations at different laser input fluences. On the example of Si, the laser melting kinetics of semiconductors was found to be noticeably different from that of metals with fcc crystal structure.

## Introduction

Short laser pulse processing of semiconductors has been rapidly progressing during last several decades. The ability to deposit large amount of energy into a tightly localized area has found a number of applications in pico- and femto-second laser machining [1,2,3] and nanostructuring [4,5] of semiconductors. Specifically, the experiments aimed on semiconductor surface modifications have revealed a particular interest in Bio- [6] and IT- technologies [7,8]. Nevertheless, while technical progress has a successful tendency of structures production on semiconductor surfaces downscale to nanometer size [9,10], the fundamental mechanisms behind such the laser-induced processes as ultrafast melting, spallation, and ablation are still a subject of active scientific discussion. While the melting time of Si on the order of 10 ns, measured in experiments [11], can be explained with a commonly accepted nucleation theory based on hydrodynamic model [12], the ultrafast solid-liquid phase transition on the order of few picoseconds and shorter implies a closer look at the kinetics of microscopic mechanism of melting at the atomic level. It was shown that under extreme conditions generated in the solids due to femtosecond laser pulse irradiation, the ultrafast melting may occur [13]. Also, significant crystal superheating in the presence of strong temperature and pressure gradients influences the melting kinetics.

Moreover, the conditions in the centre of the laser spot lead to lateral confinement of the material, so that the excited solid is subjected to the one-dimensional expansion toward the surface. This effect may influence the stability of crystal against the melting process [14].

At present, there is a number of modern approaches for the theoretical description of laser excited semiconductor solid. Details of electron dynamics and material's reaction can be studied with sophisticated kinetic methods [15,16,17,18] and *ab initio* MD approaches [19,20,21,22,23]. Larger ensembles and longer timescales are preferably studied with continuum methods. For instance, hydrodynamic models [24,25,26] allow for a good precision and are widely accepted among the experimentalists for the obtained data analysis. Due to its low computational cost, relatively simple implementation, and meanwhile ability to account for laser energy absorption, fast electron heat conduction, and strong laser-induced electron-phonon non-equilibrium, so-called Two-Temperature model (TTM) [27] has become one of the most famous continuum approaches in ultrashort laser pulse experimental data interpretation and material properties determination [28,29,30,31,32,33]. The essential problem of the continuum approaches, however, is their assumption of more or less local equilibrium conditions for a description of the laser-induced phase transition processes in the solid such as ultrafast laser melting, spallation, and ablation processes. Together with strong sensitivity of the hydrodynamic systems to variations in parameters and the usages of phase diagrams and equation of states derived in the assumption of the equilibrium conditions as well, all this limits the validity of such models' applications. This situation becomes even worse specifically under conditions realized during ultra-short (pico- and femto-second) laser pulse experiments, where the investigated material is driven to extreme transient states and non-equilibrium phase transition processes become dominant [34]. Noteworthy, that the inclusion of such effects as crystal structure and defects, as well as non-thermal phenomena, into a continuum approach for a microscopic analysis of the femtosecond laser-induced phase transition mechanisms is in general questionable.

Another possible technique to model the response of semiconductor solids to ultrashort laser irradiation is the classical Molecular Dynamics (MD). The classical MD is based on the solution of Newtonian equation of motion for every particle in the system in three-dimensional space using empirical or *ab initio* interatomic potentials. The essential advantage of the MD method is that no assumptions are made regarding the kinetics of non-equilibrium phase transitions at atomic level. The kinetics of the laser-induced processes therefore follows *a priori* from the interatomic potential only. Specifically, the MD can provide insights into the atomic-level mechanisms of non-equilibrium phase transitions. This method has been demonstrated to be an efficient tool for a microscopic analysis of the melting mechanisms under conditions of overheating in both the bulk of a crystal [35,36] and in systems with free surfaces [37,38]. Simulations of boiling, spinodal decomposition and fragmentation of a metastable liquid [39,40,41], generation of defects and propagation of laser-induced pressure waves [42,43] and laser ablation [44,45,46,47,48] have been also reported. In case of Si, the classical MD model, implementing for example such potentials as Stillinger-Weber [49] or Modified Tersoff (or MOD) [50], can describe not only the crystal structure, bulk modulus and cohesive energy, but reproduces very well such important thermophysical properties of Si material as heat capacity, equilibrium melting temperature, and volume of melting. The microscopical analysis of the atomic system in MD approach results in availability of the full statistical information and readily allows for the calculation of all macroscopic parameters such as temperature and pressure. Along with modern computational technologies and utilization of the parallel algorithms, the MD approach can cover temporal and spatial scales big enough to be directly attainable in the experiment for a direct comparison with the experiment [51]. All above, in essence, justifies the applicability of the MD method in investigation of the kinetics of the ultrashort laser-induced processes in solids and in particular in semiconductors.

The classical MD method, however, is not directly applicable for the simulation of ultrashort laser interactions with semiconductors. For instance, since the electronic contribution to the thermal conductivity of the laser-excited semiconductors is dominant, the conventional MD method, where only a lattice contribution is considered [52], significantly underestimates the total thermal conductivity. This leads to an unphysical confinement of the deposited laser energy in the surface region of the irradiated target. Moreover, the laser energy deposition by multiphoton absorption and a transient state of strong electron-lattice non-equilibrium cannot be reproduced since the electrons are not explicitly presented in the model.

The analysis of the mentioned models suggests that it is possible to consolidate different techniques within a single computational approach. Thus, a model unifying the atomistic description of the kinetics of the laser-induced phase transition processes along with the description of photo-excited free carriers dynamics in the continuum can be constructed for semiconductors (in this work Si), as it was realized in the atomistic-continuum approach (MD-TTM) for metals [53].

Several similar attempts aiming on the simulation of silicon under the conditions of laser irradiation has been recently undertaken. The TTM-like approaches [54,55,56,57] provide many insights into the reaction of the material. However, the detailed kinetics of the material modification may only be obtained with more sophisticated approaches. In [58] the authors show that the hydrodynamic effects play an important role in the description of the surface modification and ripple formation. At the same time, the kinetic approach [16], though more difficult in the implementation, leads to deeper understanding of the processes involved, such as the influence of possible local nonequilibrium in electronic subsystem on the material evolution. In ref. [59] the authors suggest the atomistic-continuum approach, in which the electronic subsystem is described with non-degenerate model based on the Boltzmann distribution of the photo-excited carriers. A pioneer work [60] presented a model for the UV laser pulse interaction with semiconductor solids on the example of Si, used a combination of MD approach (to describe the atomic subsystem) and Monte-Carlo method (for the electrons and holes kinetics, applying the assumption of free carriers). Authors accounted for the changes in the interatomic bonding due to carrier photoexcitation by omitting the attraction part of the interatomic potential. Another work reported in [61] was aimed to account for the changes in the atomic bonding of Si connected to the electron excitation. The corresponding potential depended on the electronic temperature as a parameter and its function was fit to DFT simulations of silicon in the assumption that the electrons and holes have common

Fermi-Dirac distribution. The authors admit, that they do not have rigorous connection between electronic states and the interatomic potential, which might lead to imprecise energy conservation [62].

In this work, we develop the atomistic-continuum approach for the modeling of short laser pulse interaction with free standing Si targets at fluences above the melting threshold. We begin with the description of the continuum approach based on TTM-like model. As this model includes the equation for the density *n* of electrons and holes, we will denote it nTTM [57] here and after throughout the paper. Then we give the details on the MD approach to model silicon material and explain the coupling of the MD and nTTM models into the single combined computational approach. Finally, we apply the formulated model for investigation of the kinetics of short laser pulse melting of Si films and discuss the obtained results. In particular, we compare the MD-nTTM model predictions with the results of the continuum approach nTTM alone and with the results of simulations of metal targets. The kinetics of laser melting of Si material will be studied under different fluences below and above the melting threshold.

## Description of the continuum approach nTTM

The continuum nTTM approach is based on the model by van Driel [55], in which the solid (silicon) is considered as two coupled subsystems: the phonons and the electron-hole free carriers. Due to laser pulse irradiation (in this work Ti:Sapphire laser at 800 nm wavelength), the free carriers are generated, electrons in the conduction band and holes in valence band, by one- and two-photon absorption processes. Both types of carriers are assumed to quickly equilibrate in the corresponding parabolic bands. To each of them, we apply separate Fermi-Dirac distributions with different chemical potentials, $\varphi_e$ and $\varphi_h$ for the electrons and holes respectively, but with a shared carrier density *n* and temperature $T_e$ [63]. We assume the Dember field prevents charge separation, consequently the two types of carriers move together. The model accounts for the system of three equations: continuity equation for free carrier density and two coupled energy balance equations, one for carriers and one for phonons. Owing to its similarity with an ordinary well-known TTM model [27], but with an additional equation for free carrier density *n*, here and later we will refer to this approach as nTTM model. The model parameters as well as the meaning of symbols can be found in Appendix I. The continuity equation for the density of excited free carriers can be written as [55]:

$$\frac{\partial n}{\partial t} + \nabla \cdot \vec{J} = S_n - \gamma\, n^3 + \delta(T_e)n, \qquad (1)$$

where $\vec{J}$ is the carrier density flux; first term on the right side $S_n$ describes the free carrier generation rate due to one- and two-photon absorption, second and third terms account for the processes of Auger recombination and impact ionization correspondingly.

Balance equation for the free carrier energy density *u* can be written as:

$$\frac{\partial u}{\partial t} + \nabla \cdot \vec{W} = S_u - \frac{C_{e-h}}{\tau_{ep}}(T_e - T_a), \qquad (2)$$

where $\vec{W}$ is the carrier energy flux. It takes into account the carrier thermal conductivity as well as the energy transfer due to the carrier flux. First term on the right side $S_u$ describes the energy absorption processes: free carrier energy gain, one- and two-photon absorption. The last term accounts for the electron-phonon energy exchange between free carriers with temperature $T_e$ and phonons with temperature $T_a$. This term strongly depends not only on the electron-phonon temperature difference, but also on the heat capacity of free carriers $C_{e-p}$ and electron-phonon relaxation time $\tau_{ep}$, which in general are functions of free carrier temperature and density. The energy balance equation for phonon subsystem therefore is written as follows:

$$C_a \frac{\partial T_a}{\partial t} = \nabla \cdot (k_a \nabla T_a) + \frac{C_{e-h}}{\tau_{ep}}(T_e - T_a). \tag{3}$$

The first term on the right side accounts for the phonon conduction process, which is in practice, however, due to small phonon conductivity $k_a$, is negligible and frequently omitted. The electron-phonon energy exchange term is dominant here.

Special attention must be paid here to the source terms describing both the rate of free carrier density growth and the corresponding process of their energy increase, taking place in Eq. (1) and (2) correspondingly, and given by:

$$S_n = \frac{\alpha I_{abs}(\vec{r},t)}{\hbar\omega} + \frac{\beta I_{abs}^2(\vec{r},t)}{2\hbar\omega}, \tag{4}$$

$$S_u = \alpha I_{abs}(\vec{r},t) + \beta I_{abs}^2(\vec{r},t) + \Theta n I_{abs}(\vec{r},t). \tag{5}$$

For ultrashort laser material interactions, a one-dimensional heating problem is often analyzed in case of heating spot size being much larger than the size of the thermally affected zone, analyzed in the modeling. The corresponding form of laser intensity at the surface ($z = 0$) in this case is:

$$I_{abs}(0,t) = (1 - R(T_a))\sqrt{\frac{\nu}{\pi}} \frac{\Phi_{inc}}{t_p} e^{-\nu[(t-3t_p)/t_p]^2}, \tag{6}$$

where $\Phi_{inc}$ is the incident fluence, $\nu = 4ln2$, and $R(T_a)$ is the reflectivity function. In this work, to prescribe the demanded incident fluence, the center of Gaussian pulse is shifted from the initial time $t = 0$ to 3 pulse duration times $t_p$, that in turn is defined at the full width at the half of maximum.

The spatial dependence of $I_{abs}$ can be found upon the solution of differential equation of the attenuation process:

$$\frac{dI_{abs}(z,t)}{dz} = -\alpha I_{abs}(z,t) - \beta I^2_{abs}(z,t) - \Theta n I_{abs}(z,t), \tag{7}$$

where $z$ is the depth into sample; the terms on the right side are responsible for one-, two-photon absorption, and for the free-carrier absorption processes consequently. According to this equation, the spatial attenuation of the laser strongly depends on the transient electron density and therefore may strongly change during the pulse.

Finally, the total energy density of free carriers (kinetic and potential parts) forms the comprised system of Eqs. (1)-(3) and is given by:

$$u = nE_g(n,T_e) + \frac{3}{2}nk_B T_e \left[H^{3/2}_{1/2}(\eta_e) + H^{3/2}_{1/2}(\eta_h)\right]. \tag{8}$$

where $k_B$ is the Boltzmann constant, $E_g$ is the band-gap, and $H^x_y(\eta_c)$ are the functions described in Appendix II. This appendix also includes the constitutive expressions (II. 5) and (II. 7) for $\vec{J}$ and $\vec{W}$ demanded for the system solution. Here we only show the equations needed to understand the general idea of the approach.

The system of equations (1)-(3) is written in the conservative form, which provides the exact energy conservation in case of numerical solution. Nevertheless it is not convenient to solve, since equation (2) includes both $T_e$ and $u$. One can repose the equation (2) with respect to $n$, $T_a$, and $T_e$ for a more handy numerical form. To do so, we have to note that the carrier specific heat capacity is $C_{e-h} = \partial u/\partial T_e|_n$, and using the Eq. (8) we can therefore write:

$$C_{e-h} = \frac{3}{2}nk_B\left\{H_{\frac{3}{2}}(\eta_e)+H_{\frac{3}{2}}(\eta_h)+T_e\frac{\partial \eta_e}{\partial T_e}\left[1-H_{\frac{3}{2}}(\eta_e)H_{\frac{1}{2}}^{-2}(\eta_e)\right]+T_e\frac{\partial \eta_h}{\partial T_e}\left[1-H_{\frac{3}{2}}(\eta_h)H_{\frac{1}{2}}^{-2}(\eta_h)\right]\right\}. \quad (9)$$

Substituting now $\partial u/\partial t$ in Eq. (2) we arrive at the diffusion-like equation for the temperature of electron-hole pairs:

$$C_{e-h}\frac{\partial T_e}{\partial t} = S_U - \nabla \cdot \vec{W} - \frac{C_{e-h}}{\tau_{ep}}(T_e - T_a)$$

$$-\frac{\partial n}{\partial t}\left\{E_g + \frac{3}{2}k_BT_e\left[H_{\frac{3}{2}}(\eta_e)+H_{\frac{3}{2}}(\eta_h)\right]\right\} - n\left(\frac{\partial E_g}{\partial n}\frac{\partial n}{\partial t} + \frac{\partial E_g}{\partial T_a}\frac{\partial T_a}{\partial t}\right) \quad (10)$$

$$-\frac{3}{2}k_BT_en\frac{\partial n}{\partial t}\left\{\left[1-H_{\frac{3}{2}}(\eta_e)H_{\frac{1}{2}}^{-2}(\eta_e)\right]\frac{\partial \eta_e}{\partial n}+\left[1-H_{\frac{3}{2}}(\eta_h)H_{\frac{1}{2}}^{-2}(\eta_h)\right]\frac{\partial \eta_h}{\partial n}\right\}.$$

As opposite to eq. (2), equation (10) may lead to accumulation of numerical errors in case of low precision of the derivative calculations of the last three terms, i.e. it is written in the non-conservative form. The errors may come from the last terms of eq. (10), which numerically play a role of extra sources in the diffusion equation. Nevertheless our semi implicit numerical scheme provides the precision, which is more than enough for our purposes. The details of the continuum approach described above and the derivation of the main equations can be found in [55].

Thus, from the system of equations (1), (3), and (10), we can fully determine the dynamics of $n$, $T_e$, and $T_a$ in 1D with the following initial boundary conditions, suitable for a free standing film:

$$T_a(z,0) = T_e(z,0) = 300 \text{ K},$$
$$n(z,0) = n_{eq} = 10^{16} \text{ m}^{-3}, \text{ ref. [64]},$$
$$J(0,t) = J(L,t) = 0,$$
$$W(0,t) = W(L,t) = 0, \quad (11)$$
$$k_a\frac{\partial T_a}{\partial z}(0,t) = k_a\frac{\partial T_a}{\partial z}(L,t) = 0,$$

where $L$ is the thickness of the sample.

To solve the above system, we use the finite difference grid mesh sketched in Fig. 1 (upper part). The sample is divided into cells according to the scheme, and the thermodynamic parameters were calculated in each cell. The spatial derivatives of $n$, $T_e$, $T_a$, $J$, $W$, $k_a\frac{\partial T_a}{\partial z}$, and $E_g$ at the interior points are approximated with the central difference, and those at the boundary are evaluated with the first-order approximation.

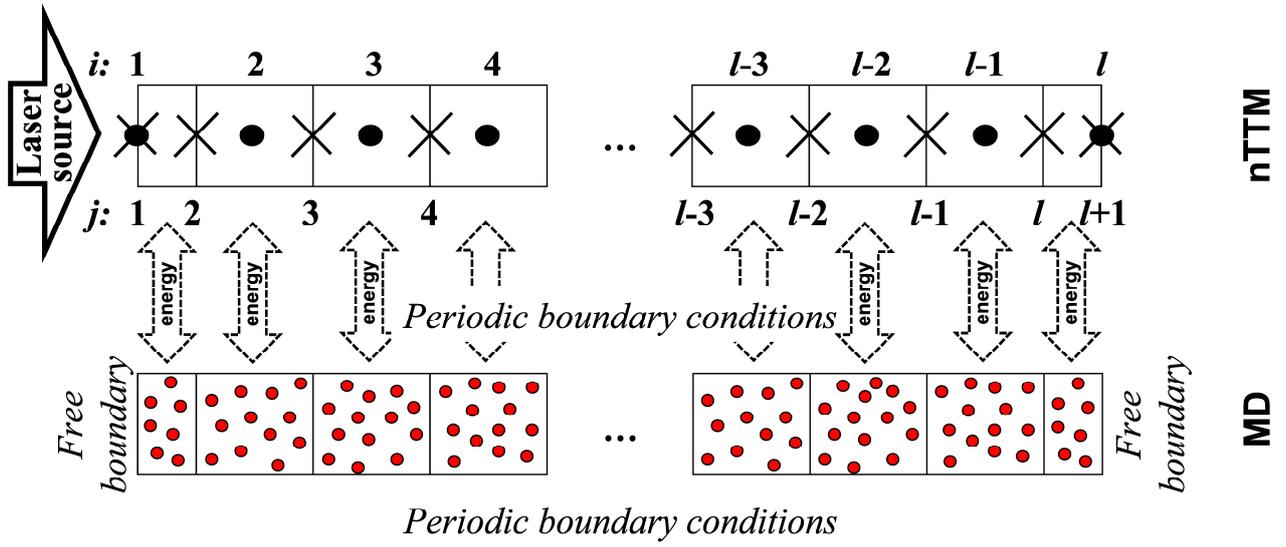

Fig. 1. The finite-difference grid mesh of nTTM model and the organization of the coupling between nTTM and MD models. Symbol "•" indicates the grid points for $n$, $T_e$, and $T_a$ ($i=1, 2, …, l$); symbol "×" indicates the grid points for $J$, $W$, $k_a \frac{\partial T_a}{\partial z}$, and $E_g$ ($j = 1, 2, …, l+1$). Red circles represent atoms.

During the solution of the considered system of equations we use semi-implicit Crank-Nicolson numerical scheme [65,66] modified for this non-linear case with the predictor-corrector algorithm. The details of the integration algorithm will be published elsewhere [67]. This approach allowed us to increase the time step of the calculations from $10^{-20}$ s up to $10^{-16}$ s as compared to the explicit scheme with an energy conservation error of less than 0.2%. Thus, the time step of $10^{-17}$ s still lets the system be solved fast enough with nearly perfect energy conservation.

When being combined with MD, the proposed implicit scheme of nTTM model solution gives a significant benefit. Since its relative calculation time is significantly shorter than that for MD part, the inclusion of nTTM in MD will not influence the calculation time of the combined model.

One of the most uncertain parameters in nTTM model is the two-photon absorption coefficient. For our wavelength (800 nm), the experimental measurements in ref. [68] yield the value of $\beta = 2$ cm/GW, while fitting to another experiment [69] gives $\beta = 55$ cm/GW. If a continuum model is involved in fitting [54], one has the value of $\beta = 9$ cm/GW. Since this parameter plays an important role in laser absorption, it noticeably influences the amount of energy absorbed in and transmitted through the sample. Our simulations show a large difference in the absorbed fluence when using different values of $\beta$ parameter. In the presented productive runs, the value of $\beta = 15$ cm/GW was chosen for calculations. This choice is justified by a good agreement between the experimental melting threshold found for 130 fs laser pulse [70,71,72] (0.26-0.27 J/cm$^2$), the one predicted from nTTM (0.29 J/cm$^2$ in our calculations), and the one obtained from the combined model (0.27 J/cm$^2$, see below). Also, in our present calculations we used a constant value of $\gamma$ for Auger recombination coefficient. However, according to [73], accounting for the effect of non-degeneracy at high carrier densities, as it is done in [54], leads to a different characteristic time of the free carrier recombination. Additionally, in [57] the authors suggested an extension to the model of van Driel, based on the Drude model. It accounts for the changes in optical parameters, namely reflectivity and free-carrier absorption coefficient, due to the highly transient free carrier density during the excitation with femtosecond laser pulses. These modifications are therefore planned for implementation in our future research.

As an example of nTTM model application, in Fig. 2 we show the electron-hole density, temperature and atomic temperature dynamics on the surface, followed by the laser pulse irradiation of 500 fs duration at the absorbed fluence of 0.0381 J/cm$^2$ (corresponding to the incident fluence of 0.15 J/cm$^2$ via the reflectivity function used in the model [74]). The Si sample thickness was taken to be 800 nm. The initial increase in the number of free carriers, followed by the laser pulse, changes to its decay due to Auger recombination and diffusion processes. Interestingly, the electron-hole temperature exhibits two elevations. The first one is associated with very low density of free carriers, which consequently have negligible heat capacity (9) and possess very little energy. The subsequent plateau corresponds to the temporal prevalence of one-photon absorption over all other processes, so that all created free electron-hole pairs fall into the same energy level $u/n = \hbar\omega$. The second increase of the temperature is connected to the free carrier absorption and its following dynamics is pretty much the same is it is for metals given by ordinary TTM [27]. Finally, the thermal energy from hot electron-hole carriers goes to the atomic subsystem of the sample leading to its gradual temperature increase upon the electron-phonon equilibration.

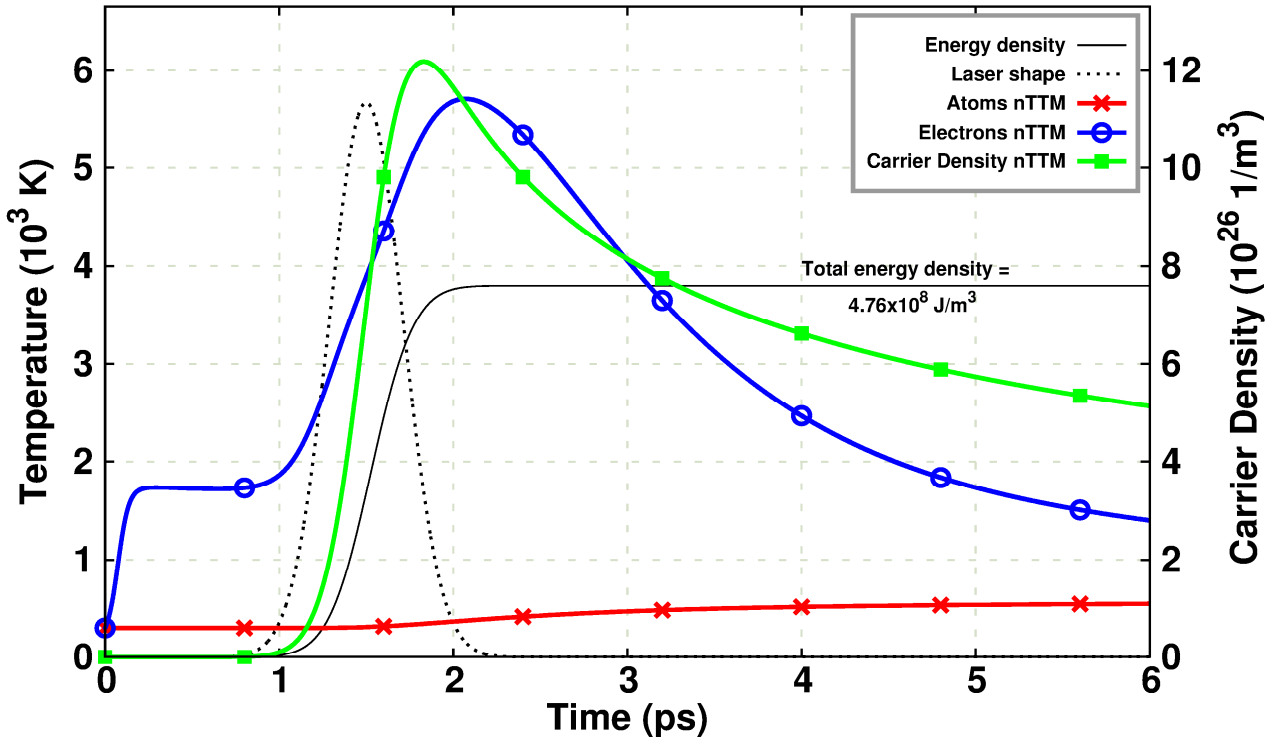

Fig. 2. (Color online) Carrier/lattice temperatures and carrier density dynamics on the sample surface according to the nTTM model followed by the 500 fs laser pulse at the absorbed fluence of 0.0381 J/cm$^2$. The sample thickness is 800 nm. The laser pulse shape and the energy density, proving energy conservation, are not in scale.

Similar simulations were also performed in [75]. The comparison of the corresponding solutions, however, shows a significant difference in the carrier's properties dynamics even for the same sample geometry and set of parameters. The essential reason for this discrepancy unfortunately originates from a major mathematical mistake during derivation of the corresponding equation for the free carrier's temperature, specifically in equations (18) and (19) of ref. [75]. As a result, the energy conservation criterion is not fulfilled there.

## MD approach

One of the main disadvantages of the nTTM model is that it is barely applicable for the description of the kinetics of considered phase transition processes in the continuum. Despite the fact that some of its problems can be solved by means of more sophisticated implementation, the induced phase transformation mechanisms, realized in the solid due to the ultrashort pulse laser irradiation, can be quite far from the equilibrium conditions. The transition times can be shorter than 1 ps and the new phase nuclei may have size of several interatomic distances only. Thus, the investigation of such processes as ultrafast melting, ablation, spallation, recrystallization, and surface effects, would require a number of assumptions to be implemented within the nTTM model. We overcome this problem by introducing the atomistic description of the solid with a classical MD method, which will eventually substitute the equation (3) in the continuum nTTM model. The MD approach is based on the solution of Newtonian equations for every atom [76]. With initial conditions (initial coordinates and velocities of all atoms), the full set of equations in 3D allows for monitoring the microscopic evolution of the system in time. This microscopic information about the system enables us to calculate any macroscopic (thermodynamic) parameters.

As we already mentioned, the choice of the interatomic potential fully determines the kinetics of laser melting and the properties of the material. For our purposes, some potentials are not suitable due to an imprecise representation of thermophysical material properties. For example, the Tersoff potential [77], although well representing the elastic properties, fails in describing the melting temperature and the volume of melting [78]. Analogously, the modified-embedded-atom method [79] is suitable for many materials, but in silicon it leads to an expansion coefficient significantly differing from the experimental value as well as to a noticeably shifted melting point [80]. Consequently, with these potentials, one can expect significantly different melting kinetics as compared with experimental results. In general, the choice of the potential is made upon performing a balance between the simplicity of implementation, computational costs and the description of important parameters. In the presented work, the interatomic interaction is described via a well-known Stillinger-Weber potential [49]:

$$V = \sum_{\substack{i,j \\ i<j}} U_2(r_{ij}) + \sum_{\substack{i,j,k \\ i<j<k}} U_3(r_{ijk}), \tag{12}$$

where $U_2$ is the two-body and $U_3$ the three-body parts. Provided that the reduced radius $r = \dfrac{r_{ij}}{\sigma}$ is less than the cutoff distance $a$,

$$U_2 = \varepsilon \times A(Br^{-p} - 1)e^{(r-a)^{-1}}, \tag{13}$$

$$U_3 = h(r_{ij}, r_{ik}, \theta_{jik}) + h(r_{ji}, r_{jk}, \theta_{ijk}) + h(r_{ki}, r_{kj}, \theta_{ikj}), \tag{14}$$

and 0 otherwise. The functions *h* are given by:

$$h(r_{ij}, r_{ik}, \theta_{jik}) = \lambda \exp(\gamma(r_{ij} - a)^{-1} + \gamma(r_{ik} - a)^{-1}) \times \left(\cos\theta_{jik} + \frac{1}{3}\right)^2, \tag{15}$$

where $\theta_{jik}$ is the angle between $\vec{r}_j$ and $\vec{r}_k$ adjoining vectors at vertex *i*, etc. This dependence on $\theta_{jik}$ leads to the crystal with 'ideal' tetrahedral angle, described with $\cos\theta_{jik} = -1/3$. Thus, the potential describes the open diamond structure of silicon solid. All the parameters in Eqs. (13) and (15), which are fitted to reproduce the properties of silicon crystal, can be found in the original paper [49].

The important thermophysical properties of the modeled silicon such as melting temperature [81], solidification and melting rates [82], thermal conductivity [83,84], bulk modulus [85], and phase diagram [60,86] reproduce the behavior of a real solid at the equilibrium conditions quite well. Moreover, the chosen potential gives better overall description of liquid phase than all frequently used potentials [87]. Appendix III presents some other important physical properties of Stillinger-Weber potential, which we will need in order to describe the melting kinetics quantitatively. However, it is known that at strong photo-excitation of semiconductors, the changes in electronic density and band structure may lead to a reconstruction of the interatomic bonding and consequently to non-thermal melting [88]. The Stillinger-Weber potential does not account for this behavior; therefore, in the combined model, we restrict ourselves to laser fluences below the non-thermal melting threshold, which was determined to be about 0.55 J/cm$^2$ for our laser parameters [54]. In our future works we are planning to introduce the modified potential, which will be able to account for the non-thermal effects as well [62].

As it was mentioned above, though MD is a powerful tool for the description of solids, it lacks the free electrons in its classical formulation, necessary for description of the laser light absorption, free carrier diffusion, fast electron heat conduction, and the electron-phonon energy exchange. For the description of free carrier subsystem dynamics of silicon solid, one can incorporate the described nTTM part into the MD method in a similar manner as it was done in [53]. In other words, we replace the equation (3) in nTTM approach with the set of Newtonian equations for each atom in the computational cell, introducing therefore the microscopic approach with all the advantages of both MD and nTTM. The combined MD-nTTM approach, therefore, will provide us with an accurate model for the description of the laser excited semiconductor solids. The details about organization of the connection between the MD and nTTM parts are described in the following section.

**The combined MD-nTTM model**

In this section we explain the organization of the coupling between the continuum and atomistic parts in the combined MD-nTTM model. The atomic subsystem in the MD part is divided into a number of volume cells, Fig. 1, lower part. Each of them corresponds to the same space geometry in the continuum nTTM part describing the electronic subsystem. In every MD cell we assume local equilibrium and calculate the temperature of atoms under assumption of equipartition between the kinetic and potential energies, based on the Virial theorem:

$$T_a = \frac{2}{3k_B} \sum_{i=1}^{N_c} \frac{m_i (\vec{v}_i^T)^2}{2}, \qquad (16)$$

where $N_c$ is the number of atoms in a given cell $c$, $\vec{v}_i^T$ corresponds to the thermal part of the atomic velocity of atom $i$ after the subtraction of the velocity of center of mass $\vec{V}_c$ of a given cell, $\vec{v}_i^T = \vec{v}_i - \vec{V}_c$. The total number of atoms $N_c$ in each cell is defined from two constraints. First, calculating the atomic temperature, we assume the applicability of thermodynamics inside each cell. It means $N_c \gg 1$. The finite value of $N_c$ leads to atomic temperature fluctuations, which decay as $\frac{1}{\sqrt{N_c}}$. Therefore, to prevent the fluctuations one needs large enough number of atoms in a cell. On the other hand, smaller number of atoms in a cell (larger number of cells) increases spatial resolution in our carrier dynamics calculations. Thus, the system is organized in such the way that $N_c$ is around 1800 (160 calculation cells), so that the local temperature is reasonably defined and at the same time a sufficient spatial resolution is ensured. In practice, however, test calculations showed no difference between $N_c$=450 and $N_c$=3600. In case of explicit continuum scheme, the choice of computational cells would also influence the time step of nTTM in a way that bigger cells lead to decreasing the computational costs.

If the material expands (shrinks) during the simulation, we add (delete) corresponding continuum computational cell. The criterion for the creation (deactivation) of a cell is that the atomic density exceeds (drops down to) the threshold of 10% of the initial average density in the system. The atoms inside the deactivated cells are included in the nearest active cell if they are located within a distance of half-cell size from the active cell. Herewith, all the physical properties of the electronic subsystem such as diffusivity and thermal conductivity, the absorption and recombination coefficients, and the electron density are scaled with the relative changes in the atomic density within each cell $c$, $\rho_c/\rho_0$, where $\rho_c$ is the current atomic density in the cell and $\rho_0$ is the initial atomic density (averaged through the whole atomic system).

Because of the relatively high computational cost of the MD method, we decrease the needed amount of material by modeling comparatively large laser spot size. This assumption enables us to use 1D diffusion model in the continuum part and suggests the MD configuration as a thin and long sample, which is hypothetically located at the center of a wide laser spot along the laser pulse propagation. Therefore one can safely apply periodic boundary conditions in the lateral sides of the sample and free boundaries at its front/rear surfaces. In this work all simulations are performed for a sample with 5 × 5 × 1472 crystal cells (294,400 atoms) with the lattice parameter of 0.54374 nm (for 300 K), which is equivalent to dimensions of 2.72 × 2.72 × 800.4 nm in X, Y, and Z axis respectively. The lateral sizes (2.72 nm) are chosen so that they are bigger than the characteristic size of liquid nuclei (~1 nm) in order to allow nuclei to grow. This allows us to describe the kinetics of homogeneous melting. The described computational set up therefore is modeling a free standing 800 nm Si film. Before the productive simulations, the sample was equilibrated at normal conditions (300 K and at $P = 0$ GPa).

The time steps of MD and continuum parts are synchronized so that $\Delta t_{MD}=k\Delta t_{nTTM}$, where $k$ is an integer number. In our case $\Delta t_{MD} = 0.5\times10^{-15}$ s and $k = 50$. This means, while the continuum part is under calculations, the MD part is waiting for the resulting energy, taken away from hot (or coming to cold) electrons accumulated over $k$ steps. As a method to account on the influence of electrons to the atomic motion we include the "friction" term into Newtonian equations of motion for every atom $i$ [53]:

$$m_i \frac{d^2\vec{r}_i}{dt^2} = \vec{F}_i + \xi\, m_i \vec{v}_i^T \qquad (17)$$

with

$$\xi = \frac{\frac{1}{k}\sum_{j=1}^{k} GV_c\left(T^j_e - T_a\right)}{\sum_i m_i \left(\vec{v}_i^T\right)^2}, \qquad (18)$$

where $G = C_{e\_h}/\tau_{ep}$ stands for the electron-phonon coupling coefficient, $V_c$ is the volume of the corresponding MD cell. The coefficient $\xi$ is calculated in the same manner as it was done in [53] based on the energy conservation law. It describes the electron-phonon interaction process so that the energy added to (or removed from) each cell of the MD system at each integration MD step would match the energy transferred between the electrons and the lattice during $k$ steps of the finite difference integration in the continuum part. This "friction approach" is chosen due to its simplicity in the implementation. We are aware of other ways of adding/removing the energy to/from the atoms (for example, Langevin thermostat [89,90,91]). Macroscopically we do not expect noticeable differences, because the friction force applied here has random directions originated from the established Maxwellian distribution in randomly oriented atomic velocities. In the following section we present the results of the combined MD-nTTM model for the investigation of short laser pulse melting of free standing Si film.

## Results and discussions

### Comparison of the nTTM and the combined MD-nTTM models predictions

In order to demonstrate the feasibility of the developed atomistic-continuum model implementation and its applicability in our research, we repeat the modeling of 500 fs laser pulse interaction with 800 nm free standing Si film with the combined MD-nTTM model at the absorbed fluence of 0.0381 J/cm². Similar to the result of the nTTM model alone, as it is reflected in Fig. 2, the lattice temperature dynamics and the temperature/density of the electron-hole free carriers are shown in Fig. 3 for the front surface cell. As one can see, the combined MD-nTTM model shows no qualitative differences with the continuum calculations. In other words, it describes the absorption of the laser light, fast electron heat conduction, free carriers diffusion, the laser-generated strong electron-phonon nonequilibrium, and at the same time contains the atomic description of the matter within the frames of a single computational approach. A minor quantitative difference in the surface temperature of atoms, ~3% (apart from the atomic temperature fluctuations, which are natural for a finite amount of atoms), between the predictions by nTTM and MD-nTTM models, we attribute to the sample expansion process and pressure dynamics, not included into the continuum calculations. This difference is not so pronounced as in case of metals [92], due to lower expansion coefficient for silicon (see Table I). The most important benefit of the combined model, however, and eventually the reward for all our efforts is that – in contrast to the continuum calculations given by nTTM approach – the combined model is now able to distinguish the whole complexity of the kinetics of laser-induced transient processes with atomic precision. This makes the MD-nTTM model a powerful tool in studying the microscopic mechanism of short laser pulse nonequilibrium phase transformations processes that will be discussed in details below. As a mean to control the accuracy of our calculations we apply the energy conservation law. For the combined atomistic-continuum MD-nTTM model the resulting error in energy conservation was found to be less than 0.45% per simulation, which we accept as a good result fulfilling our demands. The comparison of the combined MD-nTTM approach with the nTTM continuum method together with the energy conservation criterion fulfillment, therefore, confirm the accuracy of the constructed model and its applicability in our research.

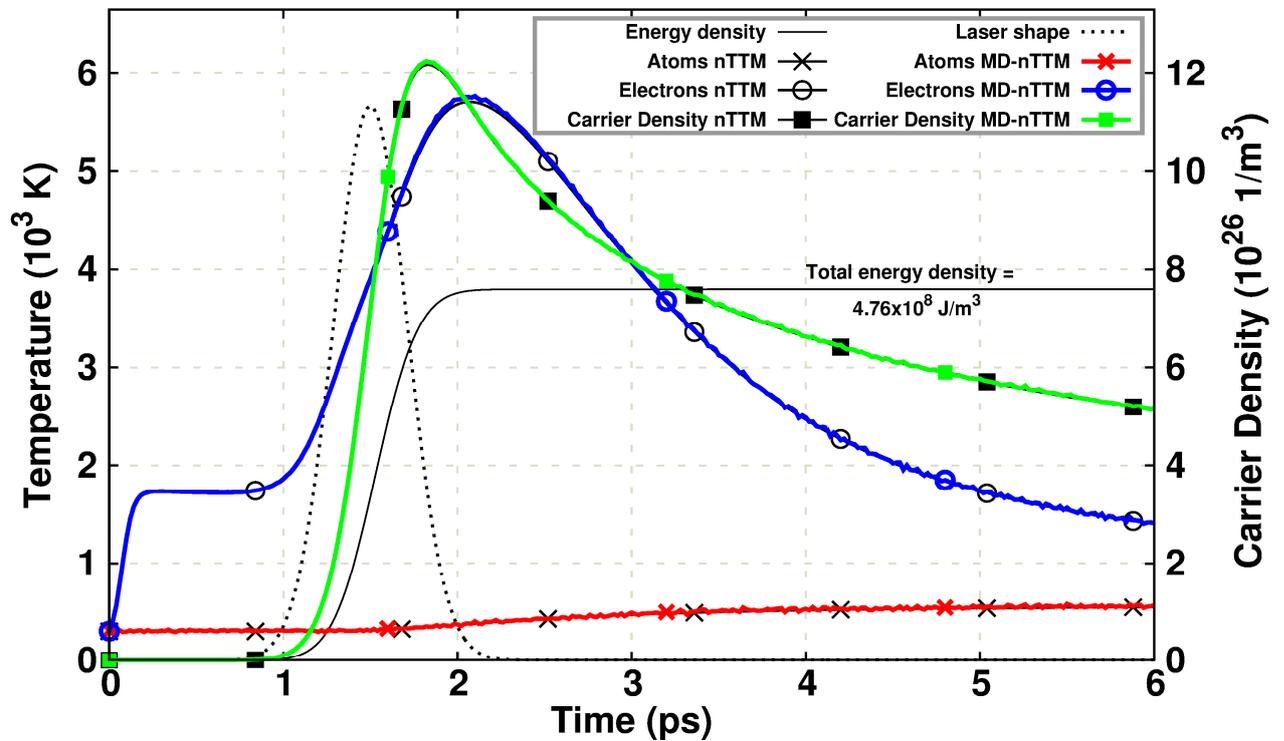

Fig. 3. (Color online) The dynamics of $n$, $T_e$, $T_a$ on the sample surface obtained with nTTM and MD-nTTM models at the same conditions as indicated in Fig. 2: 800 nm sample thickness, 0.0381 J/cm$^2$ absorbed fluence, 500 fs pulse duration. The laser pulse shape and energy conservation are shown out of scale.

Kinetics of short pulse laser melting of Si

The developed above MD-nTTM approach is applied here to study the kinetics of short laser pulse melting of free standing Si films. The parameters of the following simulation were taken according to the experiment [70]: 130 fs laser duration, 0.42 J/cm$^2$ incident fluence (0.209 J/cm$^2$ absorbed fluence given by the used reflectivity model), 800 nm laser wavelength. The sample thickness was taken to be 800 nm. From our test simulations we found that this value already represents a thick sample, since no noticeable quantitative or qualitative differences in melting kinetics are present with respect to the targets of 450 nm and 2000 nm thickness. The details of the kinetics of melting process can be extracted from the sequence of the following contour plots and atomic snapshots. Fig. 4a represents the contour plot of the percentage of molten material (calculated as the number of "molten atoms" in a cell related to the total number of atoms in it). One atom is considered as "molten" if its central symmetry parameter is lower than a threshold value of 0.9825 (see Appendix IV). Fig. 4a allows to observe the liquid nuclei generation and their growth in time. The position of the melting front was mapped by the local volumes where 50% of material is molten. The curve is shown with black solid line and, for convenience, replotted on the other contour plots, Fig. 4b and Fig. 4c, which show the contour plots of temperature and pressure evolution respectively. The white frames (rectangles) on the plot in Fig. 4a represent the places of the corresponding snapshot series, Fig. 5–Fig. 7, and are numbered correspondingly. These snapshots are taken at different moments of time after the laser pulse and at different depths. On Fig. 5–Fig. 7 every point represents an atom. Colors of atoms reflect the local structure according to the central symmetry parameter: the atoms having crystal surrounding are shown in blue color, whereas those submerged in liquid ambient are shown in red. The values of the central symmetry parameter for certain depths are shown for each particle at the figures below the atomic structure.

Upon the laser pulse absorption, the generated strong electron-phonon nonequilibrium leads to the electron-phonon energy exchange, which in turn causes the elevation of the atomic temperature to $T>T_m$ ($T_m$ = 1683K, see Appendix III) up to the depth of ~250 nm within ~4 ps (Fig. 4b; in order to provide a better view, the plot only shows 210 nm from the top of the sample). Before the onset of melting, however, the strong heating rate results in the increase of the pressure (see Fig. 4c). As discussed in Appendix III (see Fig. 11a), due to different slope of the melting curve from that in metals [93], higher pressure reduces the temperature needed to melt the crystal (by both homogeneous and heterogeneous mechanisms) and therefore speeds up the melting process. In the first picoseconds, the sample has slightly lower pressure near the surface at a higher temperature and slightly higher pressure in the depth (up to ~100nm) at a lower temperature. This results in the onset of homogeneous melting taking place simultaneously from the surface up to the depth of ~100 nm within 4 ps. Noticeably, the melting speed at this point is much higher than the speed of sound. The significant excess of thermal energy can not convert the solid into the liquid state by means of classical heterogeneous solid-liquid interface propagation, and the homogeneous melting mechanism takes a massive character, Fig. 4a, rectangle 5 and the corresponding snapshot series on Fig. 5. Such the phenomenon has been already suggested in theoretical work [94].

Here we can see a major difference in the melting process from that observed for metals [95] (which experience expansion upon melting by 3-5% [96]). In Fig. 4c, we notice that the initial laser-induced internal compressive stresses are relaxed not by the propagation of the pressure wave across the sample, but essentially by the melting process itself. From the properties of the represented material and Fig. 12a, we understand it by the fact that the melting process leads to the material volume contraction by approximately 7.5% (see Appendix III), which results in the pressure drop. Consequently, the propagating of an unloading wave decreases the pressure of remaining solid chunks of the material, which in turn reinforces their stability against the melting process. This is reflected in abruptly irregular curve of the melting front (Fig. 4a). Upon the relaxation of the tensile stresses via the propagation of the pressure waves, most of the remaining solid chunks loose their stability and the melting immediately finishes. It is reflected as the disappearance of yellow color in Fig. 4a. At the places, where the temperature is not large enough, the melting proceeds further by the classical heterogeneous mechanism at a much lower rate, which can be seen in the rectangle 6 of Fig. 4a and on the corresponding snapshots, Fig. 6. Finally, the rectangle 7 on Fig. 4a and the corresponding snapshot series on Fig. 7 indicate the areas where temperature-pressure interplay mechanism did not result in the onset of melting process neither homogeneous nor heterogeneous one. The process of new phase nuclei generation can be seen as noise consisting of small molten nuclei. None of them, however, will eventually exceed the critical size for the onset of solid-liquid transition process.

Later (around 200 ps after the laser pulse; not shown), the mentioned negative pressure wave, originated from the material shrinkage, is reflected from the rear surface and returns to the front surface as a compressive wave. This decreases the stability of the lattice and therefore temporary speeds up the front propagation. See also section "Melting depth versus fluence" and Fig. 10 for more details.

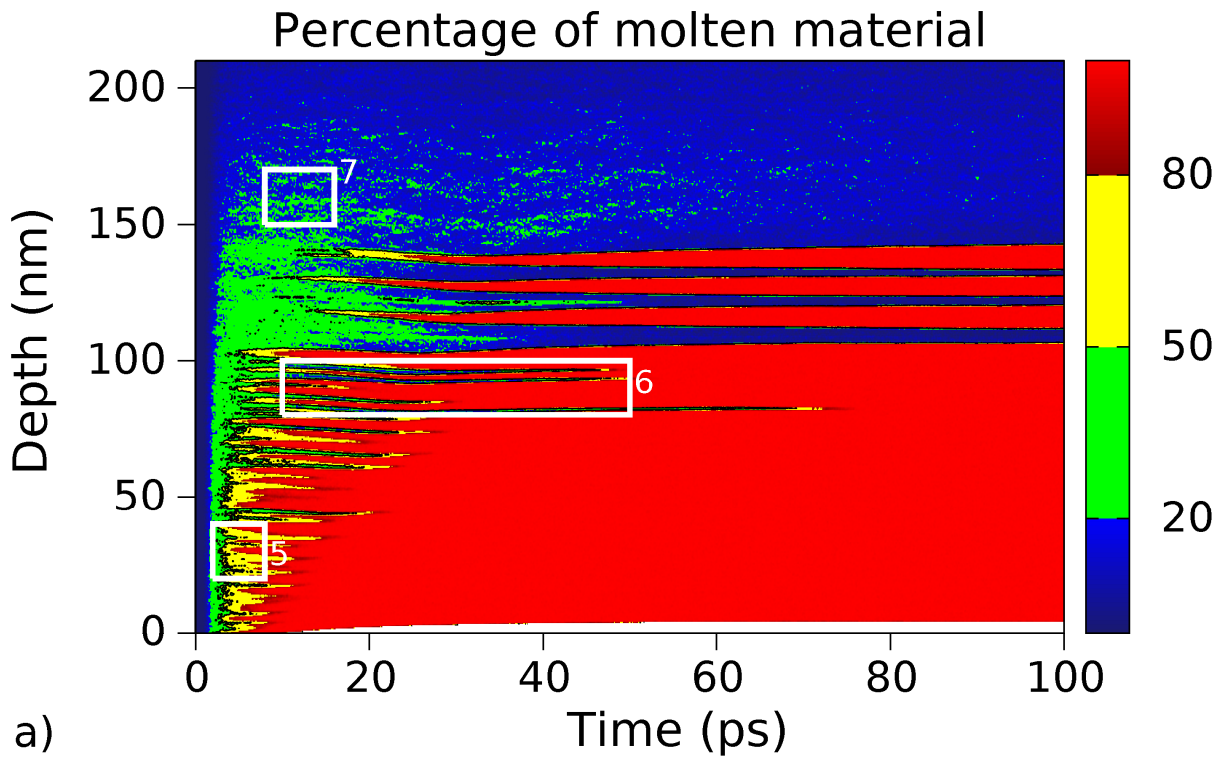
a)

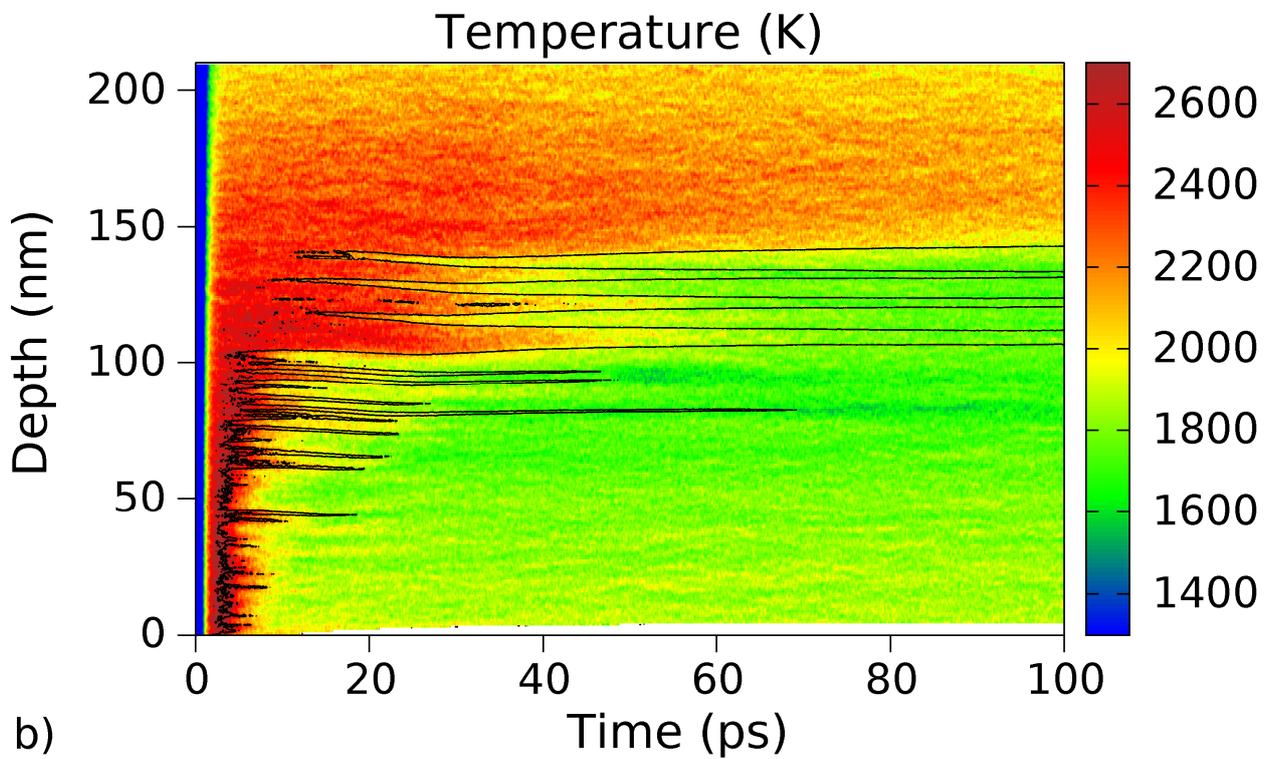

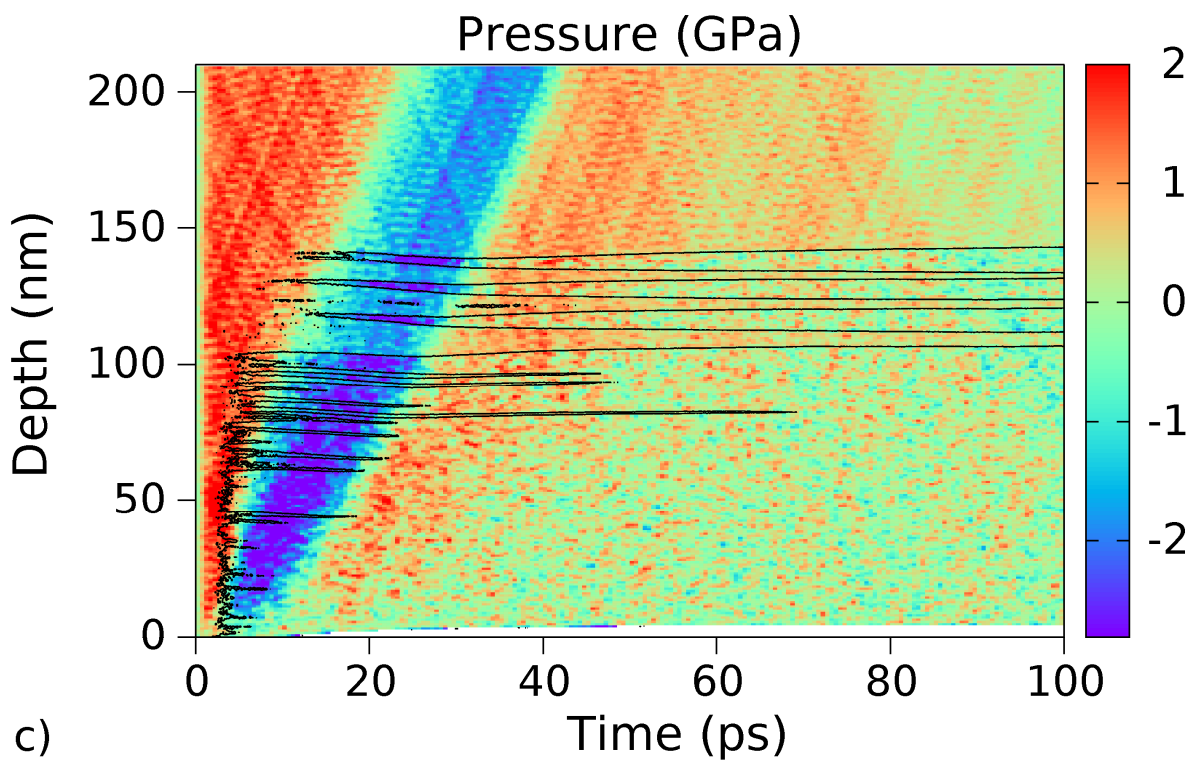

Fig. 4. (Color online) Contour plots of (a) percentage of molten material (according to the central symmetry parameter, see Appendix IV), (b) atomic temperature, (c) atomic pressure, obtained from the simulation of 130 fs laser pulse focused on 800 nm silicon film at the absorbed fluence 0.209 J/cm$^2$. The rectangles on the plot (a) show the corresponding positions of the atomic configuration snapshots presented in Fig. 5– Fig. 7. The black solid line shows the position of melting front assuming 50% of material is molten. For a closer look we show the sample depth only up to 210 nm, where the phase transition processes are taking place.

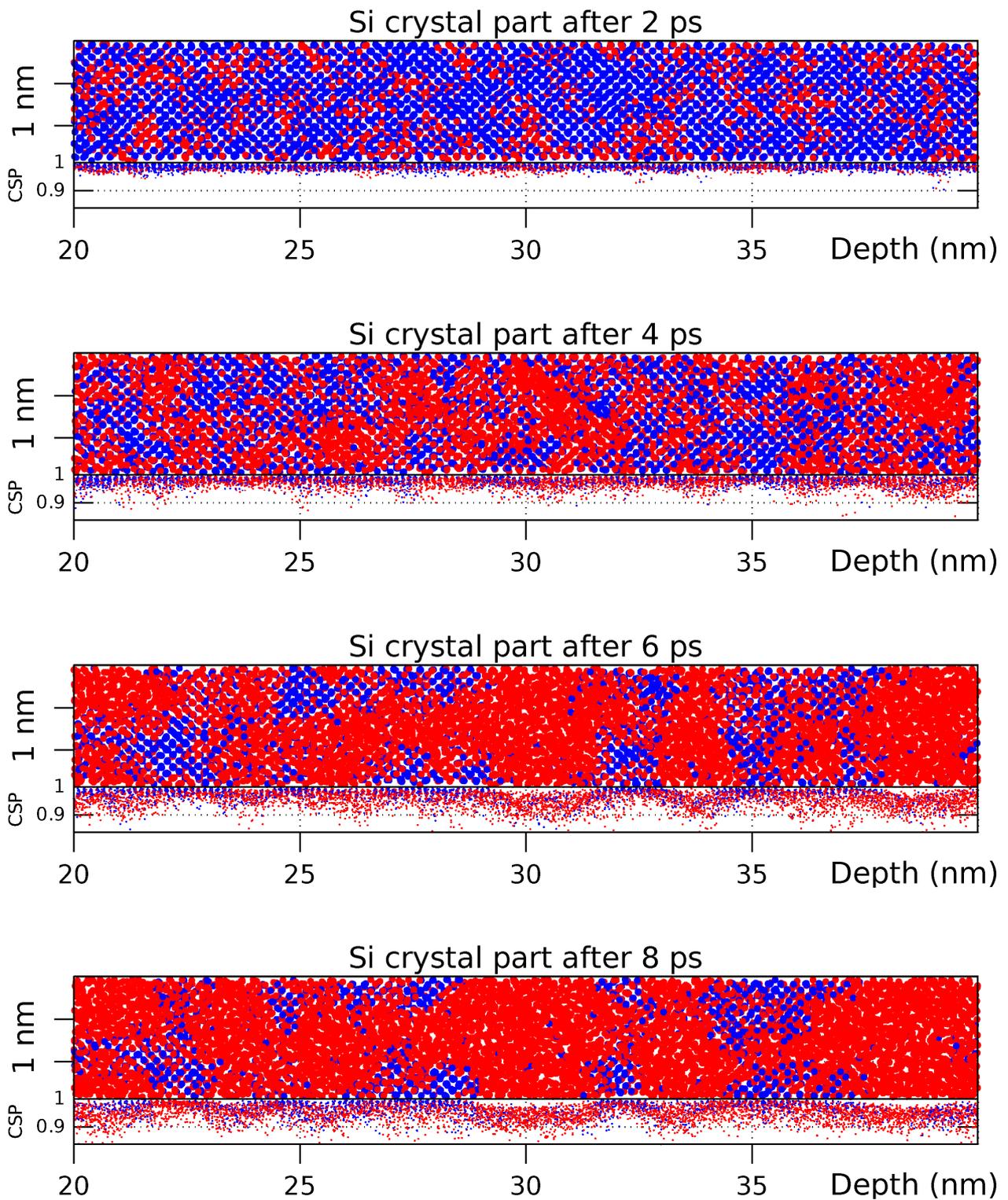

Fig. 5. (Color online) The snapshots of the simulation taken at different moments of time after the laser pulse (130 fs duration, 0.209 J/cm$^2$ absorbed fluence) at the depth of 20–40 nm (see rectangular 5 in Fig. 4a). The original laser pulse was directed from the left to the right. According to the central symmetry parameter (shown below of each snapshot), the atoms with crystalline surrounding are shown in blue color and those submerged in liquid ambient are shown in red. This series shows the presence of only the homogeneous melting mechanism at this depth.

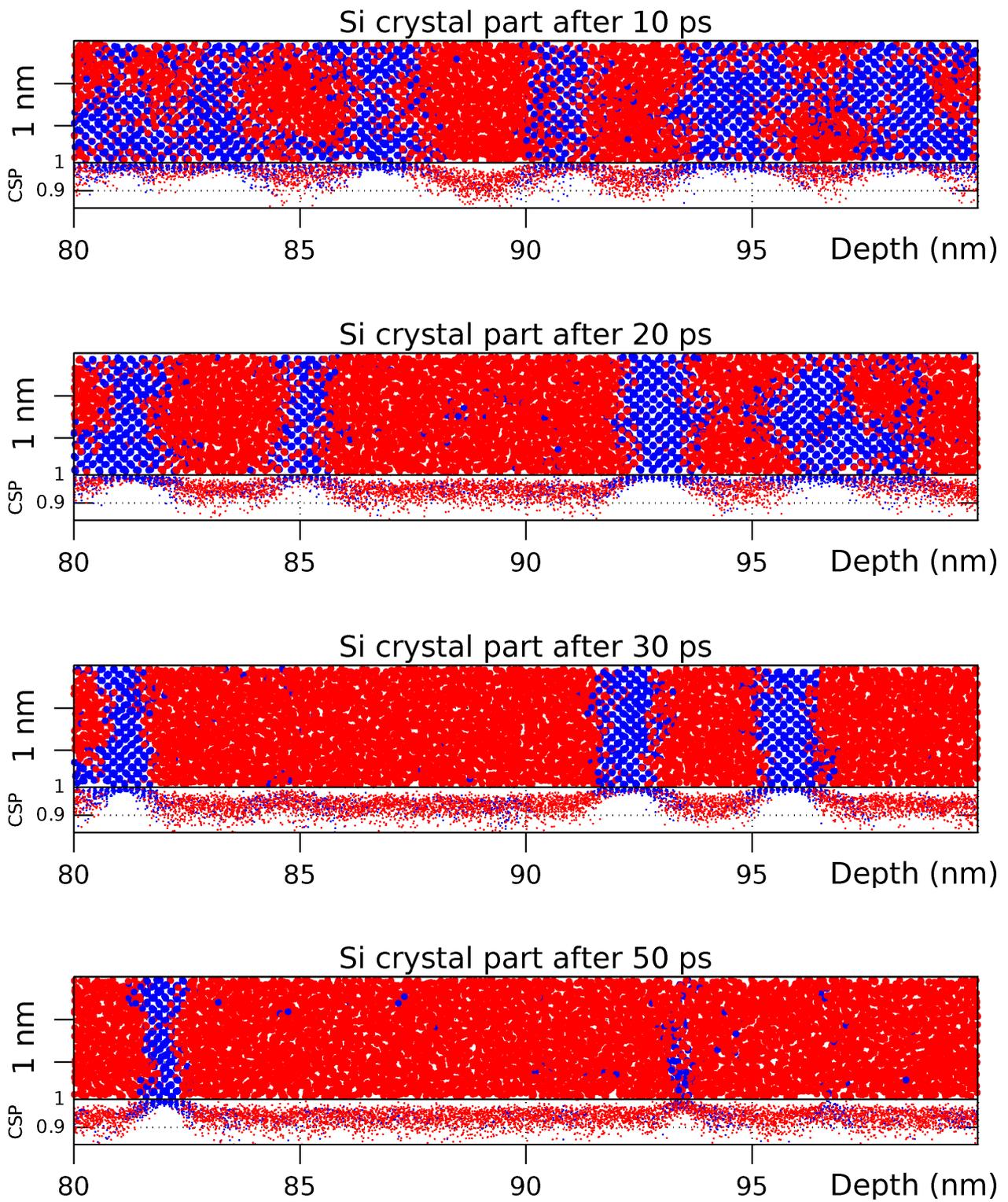

Fig. 6. (Color online) The snapshots of the simulation taken at different moments of time at the depth of 80-100 nm. This is the same simulation and style as in Fig. 5, but refers to rectangular 6 in Fig. 4a. The series shows the presence of both homogeneous and heterogeneous melting mechanisms at this depth.

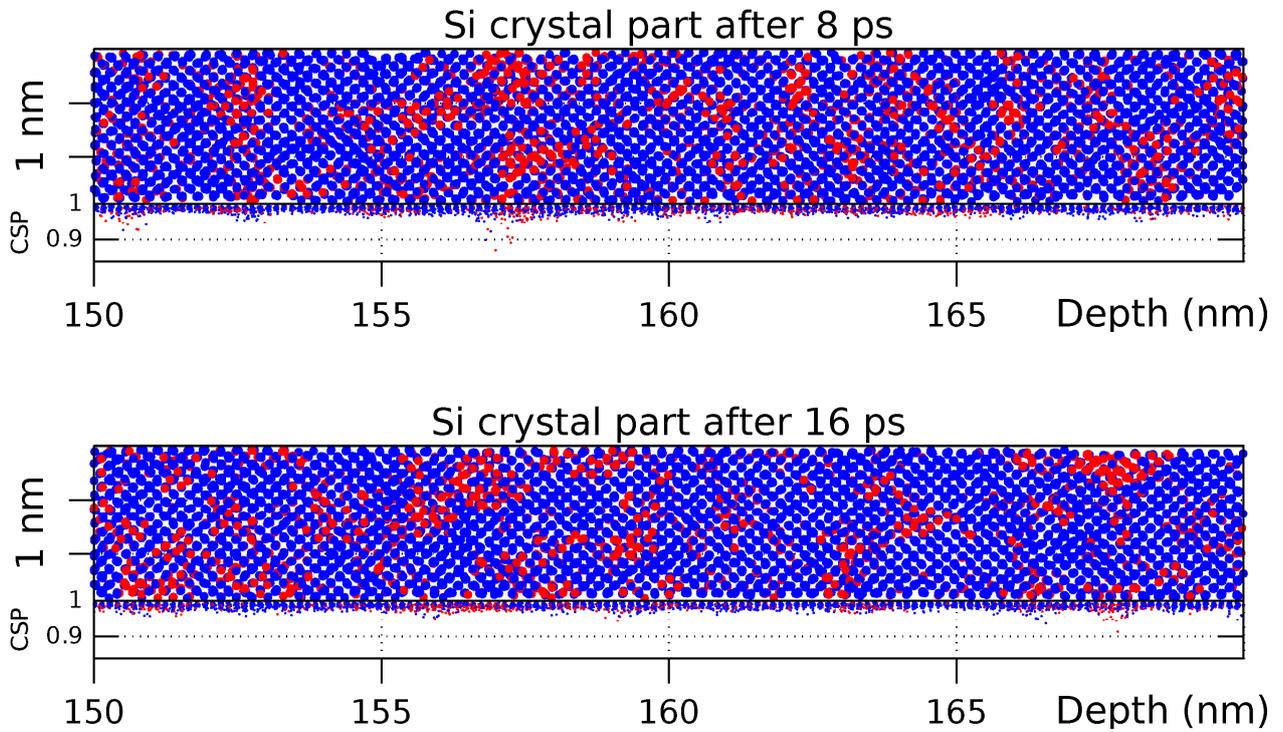

Fig. 7. (Color online) The snapshots taken at different moments of time at the depth of 150-170 nm. This is the same simulation and style as in Fig. 5, but refers to rectangular 7 in Fig. 4a. The series shows the generation of liquid nuclei that randomly appear and disappear. Under the realized conditions at this depth, however, the liquid nuclei do not grow, which preserves this part of solid from the onset of melting.

For comparison, in Fig. 8 we show the contour plots for another simulation with a smaller absorbed fluence, namely 0.161 J/cm$^2$ (corresponding to the incident fluence of 0.34 J/cm$^2$ via the reflectivity function used in the model). All other conditions in this simulation are the same as before. Fig. 8a shows the contour plot of the molten material percentage (it is shown with different scale, compared to the other contour plots, in order to provide a better view). In contrast to the result of the higher fluences simulations, in this case the melting process is fully heterogeneous. Fig. 9 shows the corresponding snapshots with well-defined solid-liquid interface. Despite the facts that the atomic temperature (Fig. 8b) rises to the melting temperature and above up to the depth of ~200 nm (as a result of the electron-phonon energy exchange) and the pressure (Fig. 8c) reaches ~1.5 GPa (as a result of thermal expansion, see the blue circles in Fig. 12a), the homogeneous melting does not occur. The reason is that the necessary overheating conditions needed for homogeneous melting do not hold long enough (~50 ps, according to the criterion suggested in Appendix III; see Fig. 11, upper line). Since the homogeneous melting process is suppressed here, it cannot prevent the development and propagation of the compressive pressure wave (Fig. 8c). The initial spatial pressure profile (at $t \approx 3$ ps) has a peak near the front surface (because the laser energy absorption is stronger near this surface), but not at the surface itself, since the surface expansion allows the pressure to relax, cf. Fig. 8a. The evolution of this pressure profile is essentially the same as in metallic targets [97,98]. It evolves as a superposition of a number of pressure sources propagating in two directions: to the front surface and to the rear surface. These waves sooner or later reach the free surfaces and are reflected with the opposite sign. Consequently, after two reflections from the front and rear free surfaces, the superposition of the wave sources gives the strongest positive pressure wave returning to the front surface (Fig. 8c).

As seen in Fig. 8a, initially the heat expansion pushes the surface away, temporarily creating 4 nm surface bump. After about $t\approx100$ ps the shrinking of the material starts to prevail and the melting front is pushed back to the frontal surface. The material continues shrinking until $t\approx200$ ps, when the mentioned returning positive pressure wave starts to provide additional energy for the heterogeneous melting mechanism near the surface. The pressure wave pushes the surface back out of material and assists the advance of the solid-liquid interface deeper into the bulk of material, in accordance with Fig. 11. For higher fluences such effect is much more pronounced and is seen as peaks in the effective melting depths (see below section "Melting depth versus fluence").

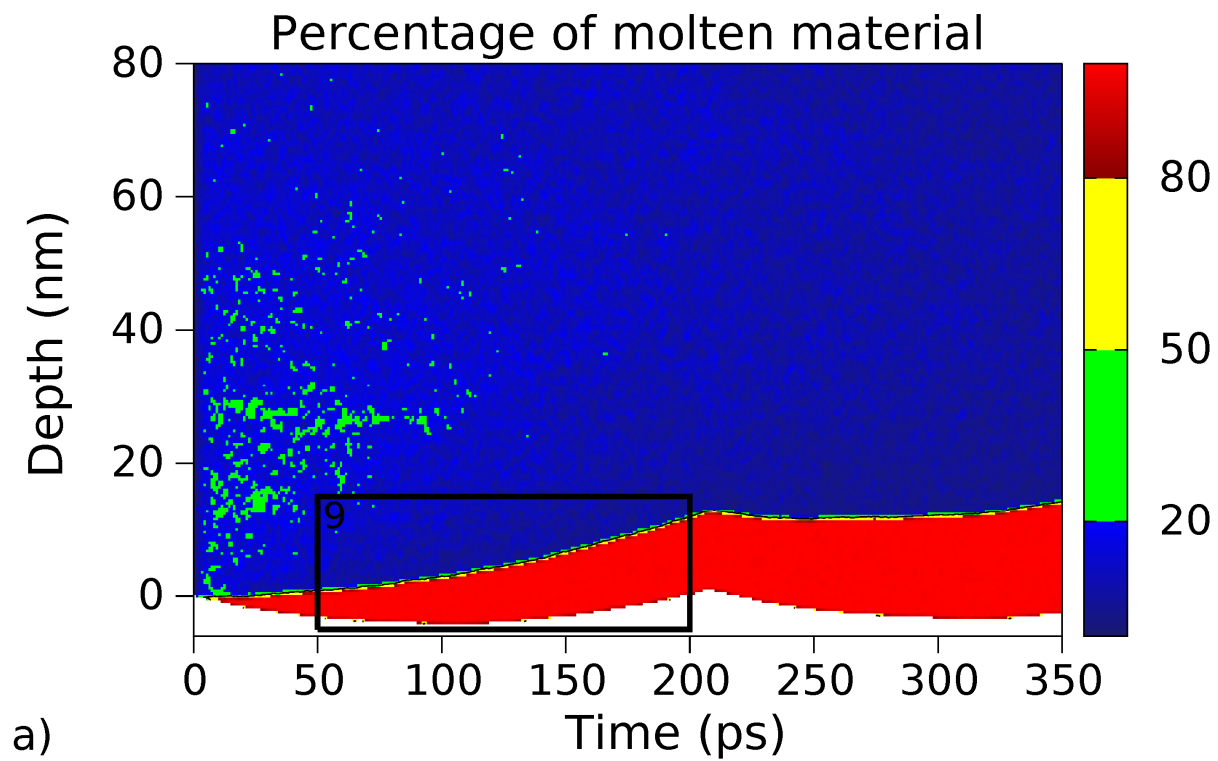

a)

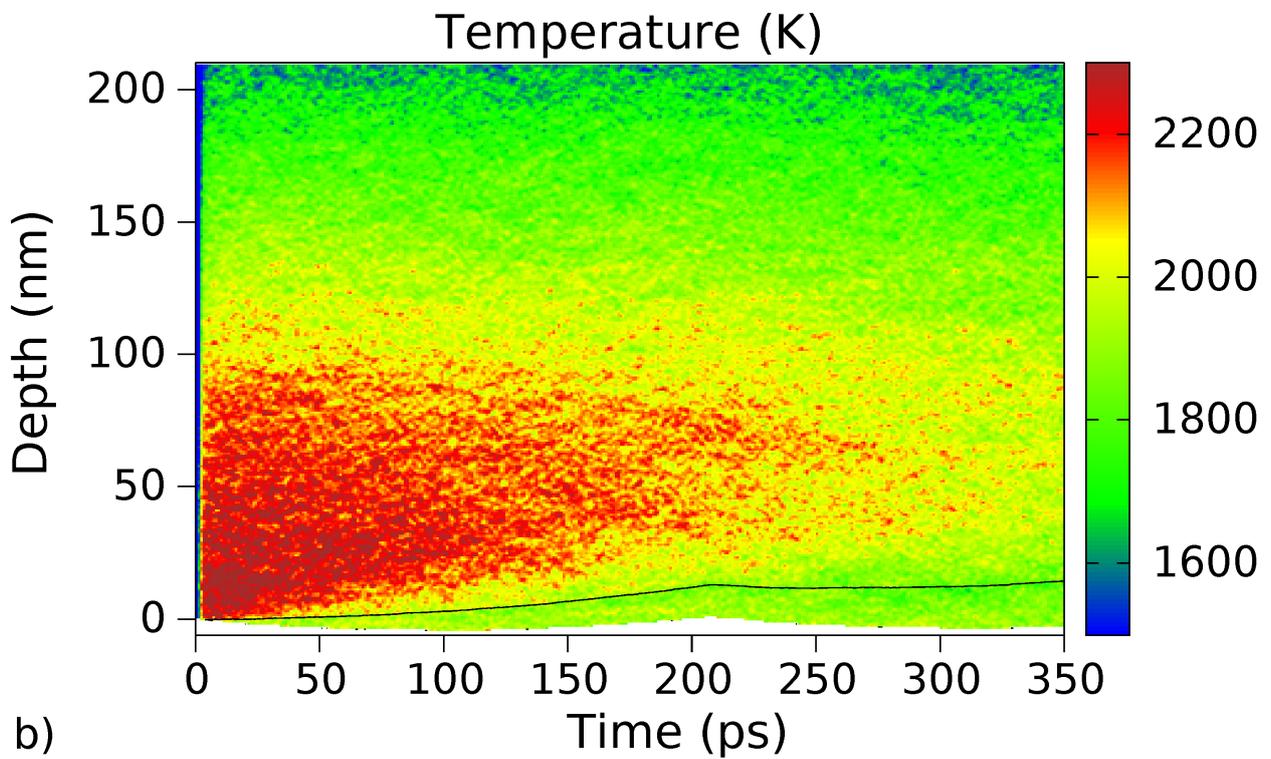

b)

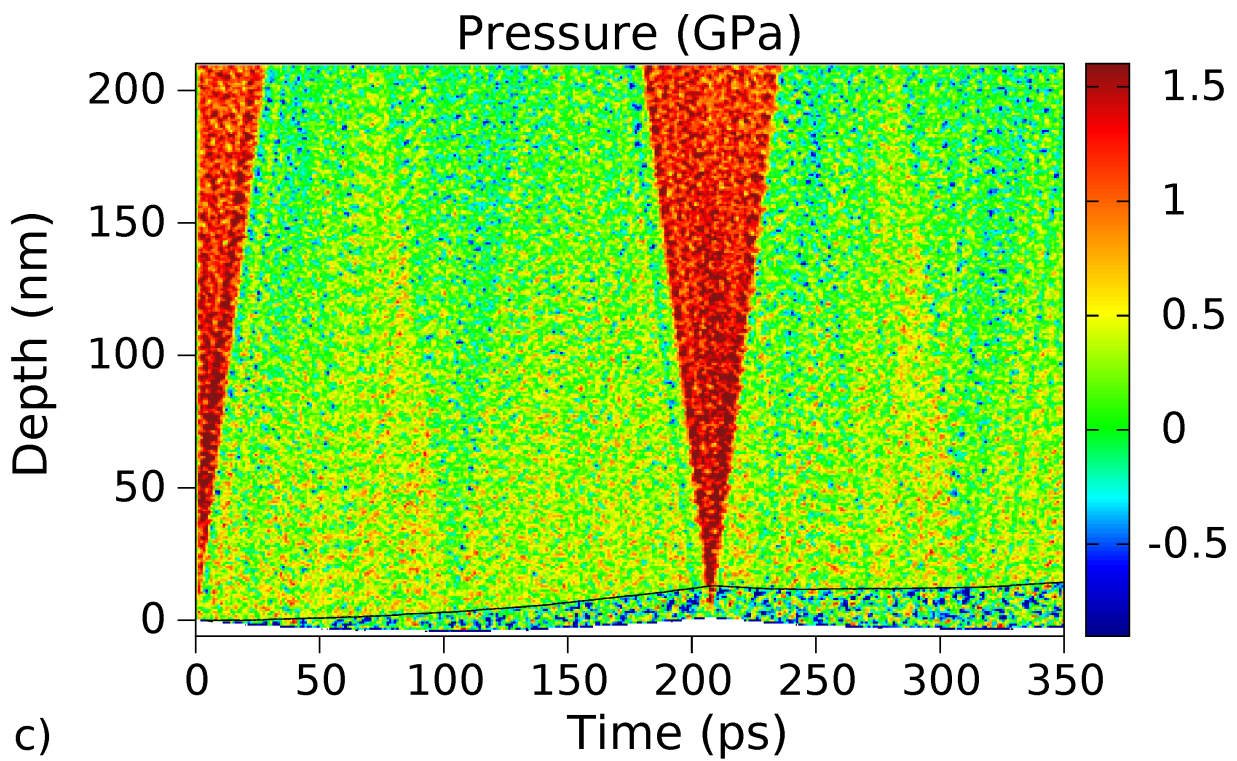

c)

Fig. 8. (Color online) Contour plots of (a) percentage of molten material (according to the central symmetry parameter, see Appendix IV), (b) atomic temperature, (c) atomic pressure, obtained from the simulation of 130 fs laser pulse focused on 800 nm silicon film at the absorbed fluence 0.161 J/cm$^2$. The plot (a) is shown with different scale in order to provide a better view of the melting front. The black solid line is obtained from the plot (a) and shows the melting front assuming 50% of material is molten. The rectangle on the plot (a) shows the position of the atomic configuration snapshots presented in the Fig. 9.

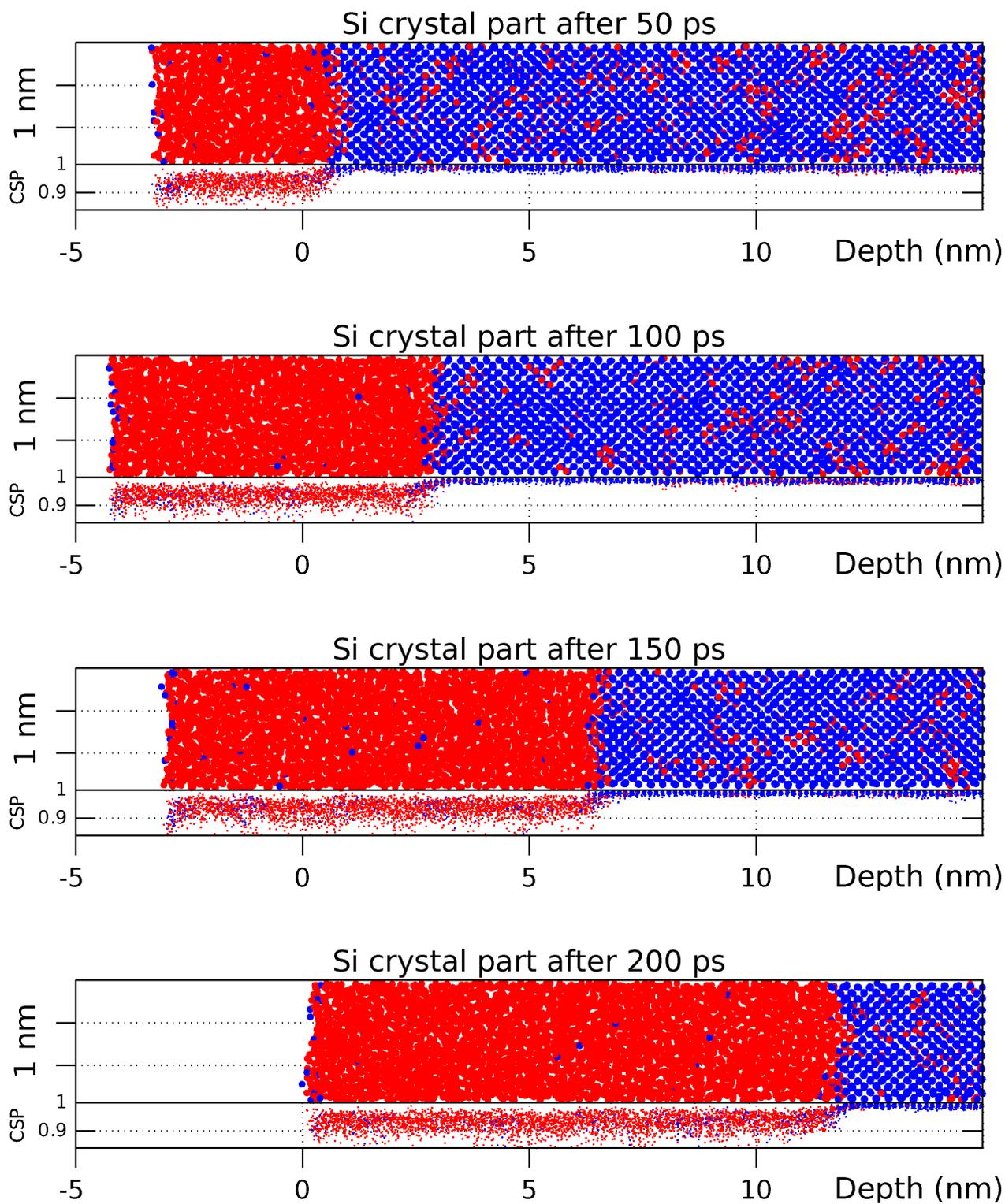

Fig. 9. (Color online) The snapshots of the simulation taken at different moments of time after the laser pulse. The laser pulse is directed from the left to the right. Only the front surface volume of the sample is shown. The original laser pulse was directed from the left to the right. The parameters of the simulation are the same as in Fig. 8. This is the same style as in Fig. 5, but refers to rectangular 9 in Fig. 8.

## Melting depth versus fluence

In this subsection, we discuss the effective melting depths for different laser fluences. The effective melting depths were found from the number of "molten atoms" (according to the central symmetry parameter, see Appendix IV) at the certain moment of time by calculating the corresponding volume of liquid for this amount (referring to the atomic density under normal conditions). Cells with <50% of molten material were excluded in order to exclude unstable liquid nuclei, such as ones in Fig. 7. We vary the total fluence, leaving all the other parameters the same: 130 fs pulse duration, 800 nm laser wavelength and 800 nm Si film thickness. The simulations were performed for the set of absorbed fluences from 0.114 $J/cm^2$ to 0.221 $J/cm^2$ (corresponding to the incident fluences from 0.26 to 0.44 $J/cm^2$). The temporal evolution of the effective melting depths is shown in Fig. 10. The model shows that for the absorbed fluences higher or equal to 0.173 $J/cm^2$, the quick initial increase in melting depth reflects that the homogeneous mechanism is dominant, whereas the following lower slope is connected to the heterogeneous process only (see Fig. 4a and Fig. 5– Fig. 7). In case of lower fluences (Fig. 10; only 0.161 $J/cm^2$ is shown) only the heterogeneous melting occurs, which is confirmed by the snapshots on Fig. 9 and the contour plot in Fig. 8a. The simulations under the same conditions, but performed for 2000 nm thick Si target confirm the threshold value of 0.173 $J/cm^2$. To our knowledge, this threshold was not mentioned in the literature before.

In Fig. 10, we observe humps in the effective melting depths for higher fluences at a time around ~200 ps after the pulse. As we already mentioned in the previous section, these sudden increases in the melting depth are connected to compression wave, incoming to the front surface and assisting the propagation of the solid-liquid interface. The maximum effective melting depths were not reached for all our simulations. For the case of absorbed fluence of 0.209 $J/cm^2$, the maximum effective melting depth was reached in 850 ps after the laser pulse and its value was found to be 143 nm. This result differs from the corresponding experimental values of 60 nm (in case of n-doped thick single-crystalline <111>-silicon [70]) and 55 nm (for 450 nm-thick undoped silicon wafers [69]). We mainly attribute it to the difference in the value of heat of fusion (see Fig. 12b and Table I), which is 31.2 kJ/mol for Stillinger-Weber potential (as is confirmed by other MD simulations using the same potential [99,100]); yet its experimental value is 50.21 kJ/mol [101]. Nevertheless, the melting kinetics is described qualitatively accurate, because of the good representation of the Clapeyron's equation at our pressure conditions and other parameters of the material (Table I). Finally, the smallest absorbed fluence of 0.120 $J/cm^2$ (corresponding to the incident fluence of 0.27 $J/cm^2$; not shown), at which the surface melting was detected, can be considered as the threshold value for the material modification and is in agreement with its experimental value [70] for our chosen value of two-photon absorption coefficient.

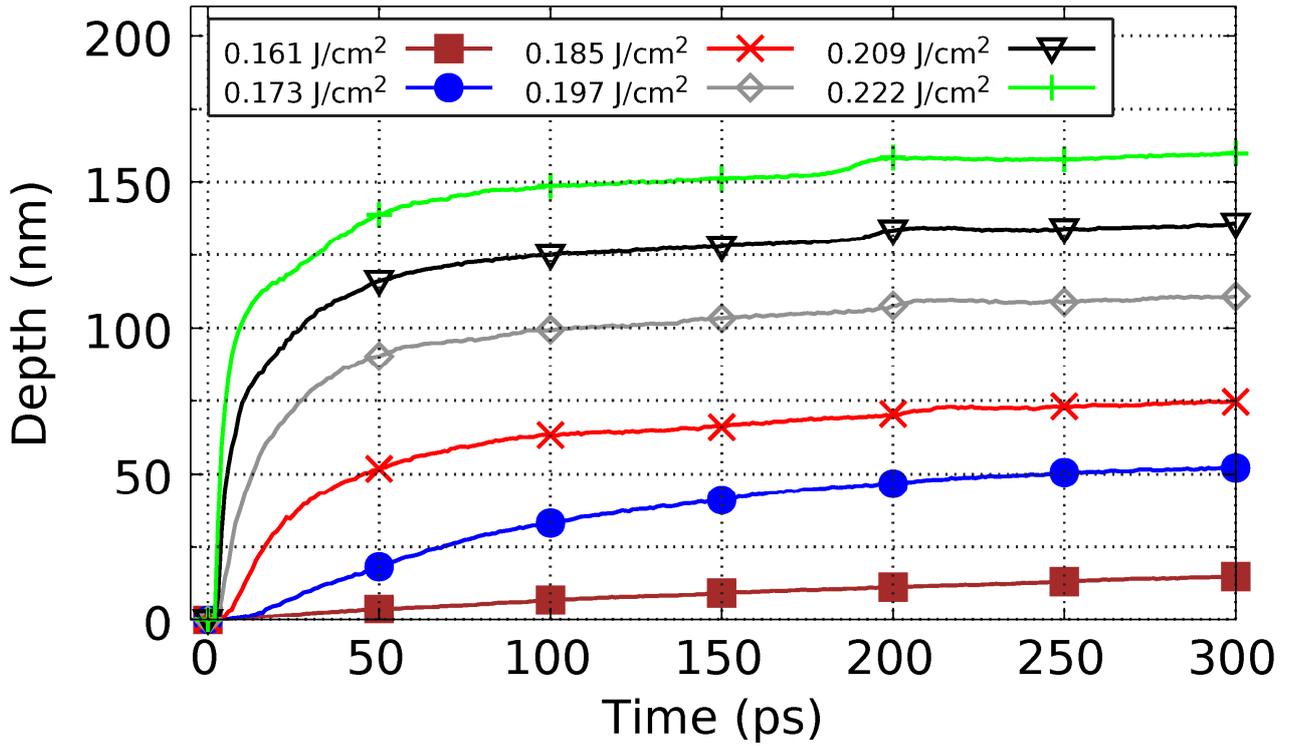

Fig. 10. (Color online) The effective melting depth depending on time for a set of the absorbed fluences: 0.161-0.222 J/cm$^2$. The lowest line corresponds to heterogeneous melting only.

**Conclusion**

We have implemented an atomistic-continuum MD-nTTM model for silicon. First calculations showed its benefits as a powerful numerical tool to study short laser pulse interaction with free standing Si films. The model is able to describe the melting kinetics with atomic resolution for fluences above the melting threshold. The approach provides microscopic insight in the processes occurring during and after the laser pulse. The energy conservation criterion fulfillment and the comparison of the obtained results with the ordinary continuum model have confirmed the reliability of the model implementation.

The laser irradiation leads to two different melting regimes: heterogeneous and homogeneous. According to the simulations, at absorbed fluences below 0.173 J/cm$^2$, the temperature and pressure – though providing the conditions for classical melting – are not high enough for the homogeneous melting. Therefore only heterogeneous melting occurs, starting from the surface, with well-defined propagation front. Higher fluences induce the homogeneous melting, which takes the dominant character, and the resulting speed of melting front propagation becomes faster than the speed of sound. The temperature and pressure interrelation allowed to explain these processes in details and revealed the consequences of negative dependence of the equilibrium melting temperature on pressure. In first picoseconds, the strong heating rate results in growing of compressive stresses, leading to positive pressure. In contrast to metals, positive pressure weakens the crystal stability against both heterogeneous and homogeneous melting, but is immediately consumed by the shrinkage of the material upon the melting process. This further results in the temporal relaxation of the material, undergone to the homogeneous melting process, and a number of liquid nuclei are incorporated into the yet crystal structure. The subsequent unloading (negative) pressure wave temporarily reinforces the stability of the residual crystalline structures. Heterogeneous process takes over the homogeneous one only after some tens of picoseconds and takes much longer time.

For fluences higher or equal to the threshold value of 0.173 J/cm$^2$, the effective melting depth rapidly increases in the first 10 ps after the laser pulse due to the quick homogeneous melting. Then the heterogeneous mechanism prevails and further melting proceeds at much slower rate. At later stages, the incoming from the rear side of the sample positive pressure waves assist the heterogeneous melting resulting in the humps in the effective melting depth. Finally, after plateaus are reached, the material starts to recrystallize back to the diamond structure.

Therefore, from the obtained results, which reproduce the experimental data values up to the limit of the precision of the interatomic potential, and from the performed microscopic analysis of the melting process, we conclude that the suggested model can be used for a number of close insights of the laser-generated non-equilibrium processes in the Si material. Moreover, with additional efforts in the implementation of the parallel algorithm and three-dimensional heat conduction, simulations can cover temporal and spatial scales big enough to be directly attainable in the experiment for a direct comparison with the experiment. Finally, based on the model construction idea and the performed analysis, one can expect that, although the analyses in this work were done for silicon, the model in general should be applicable for other semiconductor solids with the appropriate interatomic potentials and the implemented properties of the electron-hole free carrier's subsystem.

The melting kinetics of semiconductors under ultrashort laser irradiation is significantly different from that of metals with fcc crystal structure due to the differences in crystallic structure and thermophysical properties. The closed-pack structure of metallic targets leads to low-pressure assisted melting, whereas in semiconductors low pressure reinforces the crystal stability against melting. In both metals and semiconductors, liquid nuclei, formed inside the crystallic solid during homogeneous melting, reinforce the stability of surrounding crystal. However, the reasons for that are different: in metals the expansion of the material during melting by 3-5% leads to increased pressure, whereas semiconductor materials contract during melting by the order of 5-7%. Particularly, this effect is strongly pronounced for silicon (the volume of melting is 7.5%), which results in its rough melting front. The effect of heat expansion is not so pronounced in semiconductors; therefore the laser induced pressure is weaker than in metals. In metals, at the fluences around melting threshold, the laser energy deposition depth can be estimated from the properties of material and is usually a constant not larger than a few tens of nm. Consequently, the areas of both homogeneous and heterogeneous melting are near to the surface, resulting in well-defined solid-liquid interface despite possible homogeneous melting. In semiconductors however the laser energy deposition depth depends on the laser parameters and transient material state. This depth is much larger than the critical size of liquid nuclei, which is reflected in several nucleation regions remaining on different depths for a long time (sometimes until recrystallization).

The application of the model, however, is limited to fluences below those, at which changes in the interatomic potential due to electronic excitations become significant. Also, the one-dimensional model for the diffusion of free carriers and heat propagation, implemented in the current approach, is valid as long as the investigated target lateral size remains small as compared to the laser spot for particular experimental applications. Whenever this assumption is no longer valid, and the laser spot on the material surface becomes comparable with the size of the damaged region (ablation crater, generated surface feature), the suggested model must account for 3D diffusion process in its continuum part. All listed limitations however are planned to be resolved in our future investigations.

## Acknowledgements

This work was supported by the DFG grants IV 122/1 and RE1141/14. The authors acknowledge Nikita Medvedev and Anika Rämer for the helpful discussions of continuum model, Markus Nießen for the assistance in the development of the continuum solution scheme, and Nikita Shcheblanov for valuable comments on the implementation. The work was partly conducted in the Institute for Laser Technology (ILT), RWTH, Aachen, Germany, department of "Nonlinear Dynamics of Laser Processing (NLD)", in the group of Prof. Dr. Wolfgang Schulz. The authors also acknowledge the support of computational facilities of University of Kassel and Technical University of Darmstadt. Additionally, the two anonymous reviewers are acknowledged for their helpful discussions.

## Appendix I
## Model parameters

Initial carrier density: $n_0 = 10^{16}$ m$^{-3}$ [64].

Initial lattice and carrier temperature: $T_0 = 300$ K.

Lattice specific heat: $C_a = 1.978 \times 10^6 + 3.54 \times 10^2 T_a - 3.68 \times 10^6/T_a^2$, J/(m$^3$K) ($T_a$ in K) [102].

Lattice thermal conductivity: $k_a = 1.585 \times 10^5 \times T_a^{-1.23}$, W/(m·K) ($T_a$ in K) [102].

Carrier thermal conductivity: $k_e = k_h = -3.47 \times 10^{18} + 4.45 \times 10^{16} T_e$, eV/(s m K) ($T_e$ in K) [103].

Band gap: $E_g = 1.170 - 4.73 \times 10^{-4} \times T_a^2/(T_a+636K) - 1.5 \times 10^{-10} \times n^{1/3}$ if $1.170 - 4.73 \times 10^{-4} \times T_a^2/(T_a+636K) - 1.5 \times 10^{-10} \times n^{1/3} \geq 0$ and 0 otherwise, eV ($T_a$ in K, $n$ in m$^{-3}$) [104, 105].

Interband absorption (taken from 694 nm laser): $\alpha = 1.34 \times 10^5 \exp(T_a/427)$, m$^{-1}$ ($T_a$ in K) [106].

Two-photon absorption: $\beta = 15$ cm/GW (see section "Description of the continuum approach nTTM").

Reflectivity: $R = 0.329 + 5 \times 10^{-5}(T_a - 300K)$ ($T_a$ in K) [74].

Auger recombination coefficient: $\gamma = 3.8 \times 10^{-43}$, m$^6$/s [107].

Impact ionization coefficient: $\delta = 3.6 \times 10^{10} \exp(-1.5 E_g/k_B T_e)$, s$^{-1}$ [108].

Free-carrier absorption cross section: $\Theta = 2.91 \times 10^{-22} T_a/300$, m$^2$ ($T_a$ in K) [109].

Electron-phonon relaxation time: $\tau_{ep} = 0.5 \times 10^{-12}(1 + n/(2 \times 10^{27}))$, s ($n$ in m$^{-3}$) [103].

Electron effective mass: $m_e^* = 0.36 m_e$ [110].

Hole effective mass: $m_h^* = 0.81 m_e$ [110].

Mobility of electrons (taken at 1000K): $\mu_e = 0.0085$ m$^2$/(V·s) [109].

Mobility of holes (taken at 1000K): $\mu_h = 0.0019$ m$^2$/(V·s) [109].

# Appendix II

# Additional equations demanded for the carrier description

In this appendix we present the additional equations demanded to complement the system (1), (3), and (10). The full set of non-linear differential equations allows therefore the description of laser irradiation of silicon in continuum. The derivation of the following expressions can be found elsewhere [55]. The electrons and holes are assumed to have two separate Fermi-Dirac distributions with shared temperature $T_e$, but different chemical potentials $\varphi_e$ and $\varphi_h$ respectively [63]:

$$f_c(E) = \frac{1}{e^{\frac{\pm(E-\varphi_c)}{k_B T_e}} + 1}, \qquad \text{(II. 1)}$$

where subscript $c$ stands as $e$ for electrons and $h$ for holes; the + sign is associated with electrons and the − sign with holes. The reduced chemical potentials are defined as follows:

$$\eta_e = \frac{\varphi_e - E_C}{k_B T_e} \text{ and } \eta_h = \frac{E_V - \varphi_h}{kT_h}, \qquad \text{(II. 2)}$$

where $E_C$ and $E_V$ are the conduction and valence band energy levels respectively, so the energy gap is $E_g = E_C - E_V$. The integration of the carrier distribution functions over the energy leads to the expressions for the carrier density (parabolic bands are assumed):

$$n_c = 2\left[\frac{m_c^* k_B T_e}{2\pi\hbar^2}\right]^{\frac{3}{2}} F_{\frac{1}{2}}(\eta_c). \qquad \text{(II. 3)}$$

The Fermi-Dirac integral is defined as:

$$F_\xi(\eta_c) = \frac{1}{\Gamma(\xi+1)} \int_0^\infty \frac{x^\xi}{1+\exp(x-\eta_c)} dx. \qquad \text{(II. 4)}$$

The electrons and holes are assumed to move together due to the Dember field preventing the charge separation. The carrier current is therefore:

$$\vec{J} = -D\left\{\nabla n + \frac{n}{k_B T_e}\left[H^{\frac{1}{2}}_{-\frac{1}{2}}(\eta_e) + H^{\frac{1}{2}}_{-\frac{1}{2}}(\eta_h)\right]^{-1} \nabla E_g + \frac{n}{T_e}\left[2\frac{H^1_0(\eta_e)+H^1_0(\eta_h)}{H^{\frac{1}{2}}_{-\frac{1}{2}}(\eta_e)+H^{\frac{1}{2}}_{-\frac{1}{2}}(\eta_h)} - \frac{3}{2}\right]\nabla T_e\right\}, \qquad \text{(II. 5)}$$

where $H^\xi_\zeta(\eta_c) \equiv F_\xi(\eta_c)/F_\zeta(\eta_c)$ and the ambipolar diffusion coefficient is:

$$D = \frac{k_b T_e}{q_e} \frac{\mu_e \mu_h H^0_{\frac{1}{2}}(\eta_e) H^0_{\frac{1}{2}}(\eta_h)}{\mu_e H^0_{\frac{1}{2}}(\eta_e) + \mu_h H^0_{\frac{1}{2}}(\eta_h)}\left[H^{\frac{1}{2}}_{-\frac{1}{2}}(\eta_e) + H^{\frac{1}{2}}_{-\frac{1}{2}}(\eta_h)\right]. \qquad \text{(II. 6)}$$

Ambipolar energy flow is the sum of diffusion and thermal energy currents inside the carrier subsystem, which can be written as:

$$\vec{W} = \{E_g + 2k_B T_e[H^1_0(\eta_e) + H^1_0(\eta_h)]\}\vec{J} - (k_e + k_h)\nabla T_e. \qquad \text{(II. 7)}$$

## Appendix III

## Properties of the Stillinger-Weber Silicon

In order to have our atomistic-continuum MD-nTTM modeling results analyzed quantitatively, we must first find out the properties of the material represented with Stillinger-Weber potential for Si [49]. Among the properties that are of essential interest for us, one can point out the equilibrium melting temperature, volume of melting, enthalpy of fusion (latent heat of melting), linear expansion coefficient, and heat capacity. In this Appendix we perform classical MD calculations in order to determine the material properties.

One of the most important parameters in our work, the equilibrium melting point, can be found from a sequence of liquid-crystal coexistence simulations. For this purpose a sample with $6 \times 6 \times 40$ lattice cells in X, Y, and Z dimensions, containing 11500 atoms, was prepared partially molten at a certain pressure. Thereafter, the sample was equilibrated over a nanosecond so that the liquid and solid phases coexisted together at the stable pressure and temperature across the whole sample. This method excludes the presence of the nucleation barrier upon the phase transformation, and we therefore surely measure the equilibrium melting temperature at the given pressure. Performing a series of such simulations at different pressures, one can apply a linear fit to the obtained data points according to Clapeyron's equation,

$$\left(\frac{dT}{dP}\right)_{T_m} = \frac{\Delta V_m}{\Delta S_m} = \frac{T_m \Delta V_m}{\Delta H_m}. \quad \text{(III. 1)}$$

This equation not only helps to define the equilibrium melting temperature for normal conditions, $T_m = 1683 \pm 2$ K (P = 0 GPa), but is also linked to other thermo-physical properties such as entropy of fusion $\Delta S_m$, volume of melting $\Delta V_m$, and the latent heat of melting $\Delta H_m$ (enthalpy of fusion). From Fig. 11 one can see the relation between equilibrium melting temperature and pressure is different from the one in metals [53,93]. Higher (lower) pressure assists (hinders) the melting process, therefore decreasing (increasing) the necessary temperature needed to melt the crystal. This result is in agreement with the silicon shrinkage during melting (see below Fig. 12a).

Fig. 11 further shows data at higher temperatures, which refer to maximum possible overheating of the crystal before the onset of homogeneous nucleation of the liquid phase (blue pluses and blue dashed line) for different pressure conditions. The corresponding simulations involved $12 \times 12 \times 12$ lattice cells with periodic boundary conditions. The crystal was considered stable in case liquid nuclei of the critical size do not appear within 50 ps after giving the velocities to the atoms. Note, that the choice of this time slightly influences the resulting maximum possible overheating temperature.

Similarly, we calculated the maximum possible overheating temperatures in case of one-dimensional expansion/contraction of the solid (red circles in Fig. 11). The calculations show that, in contrast to metals [53], the presence of lateral crystal confinement for the realized heating conditions does not noticeably affect the maximum possible overheating of the crystal from that obtained at the homogeneous conditions. The latter therefore can be used in our analysis.

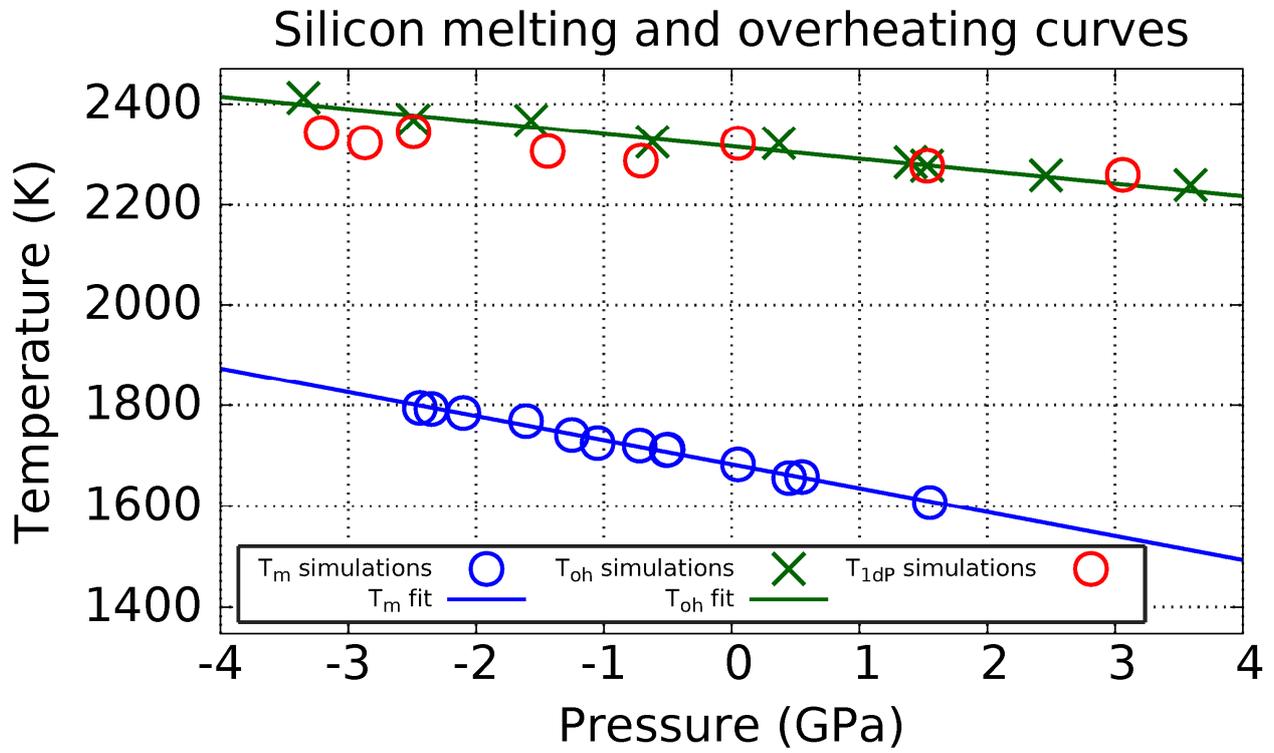

Fig. 11. (Color online) Equilibrium melting temperatures and maximum overheating temperatures for different pressure conditions. Green crosses are the results of liquid-crystal coexistence simulations with MD. Blue pluses are the results of MD simulations on the maximum overheating temperature, which crystal can have without melting within at least 50 ps (see text). Red circles were calculated at conditions similar with blue pluses, except for the expansion and contraction of the sample were one-dimensional. Solid and dashed lines are the results of linear fitting procedures.

Other thermophysical properties such as volume of melting $\Delta V_m$, latent heat of melting $\Delta H_m$, linear expansion coefficient, and heat capacity of the represented material can be found from another sequence of constant pressure / constant temperature simulations. In these simulations for a sample with dimensions of 8x8x8 lattice parameters, consisting of 4096 atoms, we can measure the volume and energy per atom as a function of temperature at $P = 0$ GPa, first for solid and then for liquid phases, Fig. 12a,b. The jump between solid and liquid curves at the melting temperature corresponds to the volume of melting, Fig. 12a, and enthalpy of fusion, Fig. 12b. Moreover, applying a linear fit to the lattice parameter as a function of temperature, Fig. 12a, and a fit of type $E = AT_a + BT_a^2/2$ for the energy as a function of temperature, Fig. 12b, we can take the derivative with respect to temperature and obtain the linear expansion coefficient and the heat capacity at zero pressure (see Table I).

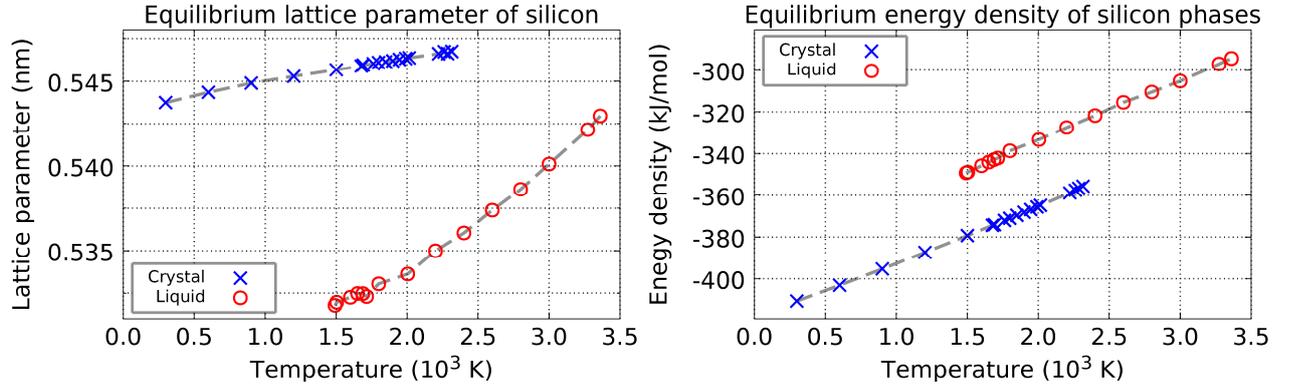

Fig. 12. (Color online) Thermophysical properties of the Stillinger-Weber silicon: (a) Equilibrium lattice parameter for different temperatures at zero pressure, (b) Equilibrium energy density for different temperatures at zero pressure. Blue circles (showing solid phase) and red triangles (showing liquid phase) are the results of MD simulations. Dashed lines are guides to the eye.

The obtained results are collected in Table I and compared with their experimental values. We can see that the material properties represented by the Stillinger-Weber potential [49] for Si generally have a good match with the experimental data. We therefore analyze our results obtained with MD-nTTM model quantitatively for the material represented with the given potential, and have some possibilities for their semiquantitative comparison with the experimental data.

Table I. Comparison of the properties between the Stillinger-Weber silicon and experimental crystal.

| Parameter | Calculated value | Experimental value | References |
|---|---|---|---|
| Lattice parameter (300 K) | 0.54374 nm | 0.54305-0.54307 nm | [101,111] |
| Melting temperature | 1683 ± 2 K | 1687-1688 K | [101,111] |
| Volume shrinkage during melting | 7.5% | 7-9.6% | [111,112,113,114] |
| Enthalpy of fusion | 31.2 kJ/mol | 50.21 kJ/mol | [101] |
| Linear expansion coefficient | $3 \times 10^{-6}$ 1/K | $2.6 \times 10^{-6}$ 1/K (300K) | [101,115] |
| Heat capacity with constant pressure | $1.94 \times 10^6 + 2.32 \times 10^3 T_a$, J/(m$^3$K) | See plotted data in the Ref. | [116] |
| Heat capacity with constant volume | $2.00 \times 10^6 + 1.84 \times 10^2 T_a$, J/(m$^3$K) | $1.978 \times 10^6 + 3.54 \times 10^2 T_a - 3.68 \times 10^6/T_a^2$, J/(m$^3$K) | [102] |
| Melting curve slope | -46.49 K/GPa | -58.7 K/GPa (near $P$=0) or -60 K/GPa | [117] and Ref. therein; [111] |

# Appendix IV

## Parameters for distinguishing between crystal and liquid structures

We use two different order parameters to distinguish between the crystal and liquid states of matter: local order parameter (LOP) [118] and modified central symmetry parameter (CSP), similar to the one in [119]. We construct them for the diamond crystal structure and average their values over the nearest neighbors (first or second neighbor shell) to prevent large fluctuations.

To calculate the LOP for atom $i$ we use the following definition:

$$LOP_i = \left| \frac{1}{6} \frac{1}{N_1} \sum_{j=1}^{N_1} \sum_{k=1}^{6} \exp(i \vec{q}_k \vec{r}_{ij}) \right|^2 . \qquad (IV.\ 1)$$

Index $j$ runs through all the neighbors of atom $i$ inside the second neighbor shell ($N_1 = 12$ for ideal diamond structure). The vectors $\{\vec{q}_k\}_{k=1}^{6}$ are defined as follows: $\vec{q}_1 = \frac{8\pi}{a}\{1;-1;1\}$, $\vec{q}_2 = \frac{8\pi}{a}\{-1;1;1\}$, $\vec{q}_3 = \frac{8\pi}{a}\{1;1;-1\}$, $\vec{q}_4 = \frac{8\pi}{a}\{0;2;2\}$, $\vec{q}_5 = \frac{8\pi}{a}\{2;2;0\}$, $\vec{q}_6 = \frac{8\pi}{a}\{2;0;2\}$. $a=a(T)$ is the lattice parameter of the crystal depending on atomic temperature (see Fig. 12a).

CSP of atom $i$ is calculated from the following expression:

$$CSP_i = 1 - \frac{\left[ \sum_{j=1}^{N_2} (\vec{r}_{ij}) \right]^2}{N_2 \sum_{j=1}^{N_2} \vec{r}_{ij}^{\ 2}} , \qquad (IV.\ 2)$$

where $j$ runs over the nearest neighbors inside the first neighbor shell $r < \frac{1}{2}\left(\frac{\sqrt{3}}{4} + \frac{\sqrt{2}}{2}\right)a$, $N_2 = 4$ for ideal diamond structure.

In case of ideal crystal, the values of LOP and CSP are 1 for every atom and when the melting occurs, they quickly decay. Both LOP and CSP allow for a sharp distinction between the two phases of silicon with the chosen criteria (0.0155 for LOP and 0.9825 for CSP). CSP has somewhat higher precision (weaker noise) and does not depend on the crystal orientation. It also allows to distinct homogeneous recrystallization inside the liquid phase. For the discussions and interpretations, the CSP is used throughout the paper. We only apply LOP for coloring the graphs of CSP below the snapshots in Figs. 5-7 and 9.

The mentioned threshold values were found from the contour plot in Fig. 4a. Initially the plot was calculated using LOP. We were decreasing the threshold for LOP starting from 1 until the first signs of (yellow) noise appear in liquid (red) area in the plot. It means the noise became balanced between the crystal (blue) area and liquid (red) area and is not bigger than 20%. This led to the threshold value of 0.0155 for LOP. On the other hand, CSP does not show any noticeable noise in Fig. 4a even when its threshold is not well adjusted. We therefore found the latter from the best match of 50% melting fronts (black curves) between LOP and CSP variants of Fig. 4a. As a result of using CSP with found threshold value, the noise in the Fig. 4a has been significantly decreased.

# References

[1] X. Liu, D. Du, and G. Mourou, Laser ablation and micromachining with ultrashort laser pulses, *Quantum Electronics*, *IEEE Journal of* **33**, 1716 (1997).

[2] B.N. Chichkov, C. Momma, S. Nolte, F. von Alvensleben, A. Tünnermann, Femtosecond, picosecond and nanosecond laser ablation of solids, *Appl. Phys. A* **63**, 109 (1996).

[3] R.R. Gattass and E. Mazur, Femtosecond laser micromachining in transparent materials, *Nature Photonics* **2**, 219 (2008).

[4] S.Y. Chou, C. Keimel, and J. Gu, Ultrafast and direct imprint of nanostructures in silicon, *Nature* **417**, 835 (2002).

[5] R. Le Harzic, D. Dorr, D. Sauer, N. Neumeier, M. Epple, H. Zimmerman and F. Stracke, Generation of high spatial frequency ripples on silicon under ultrashort laser pulses irradiation, *Phys. Proc.* **12**, 29 (2011).

[6] E. Stratakis, A. Ranella and C. Fotakis, Biomimetic micro/nanostructured functional surfaces for microfluidic and tissue engineering applications, *Biomicrofluidics* **5**, 013411 (2011).

[7] A. Mathis, F. Courvoisier, L. Froehly, L. Furfaro, M. Jacquot, P.A. Lacourt, and J.M. Dudley, Micromachining along a curve: Femtosecond laser micromachining of curved profiles in diamond and silicon using accelerating beams, *Appl. Phys. Lett.* **101**, 071110 (2012).

[8] M.K. Bhuyan, F. Courvoisier, P.A. Lacourt, M. Jacquot, R. Salut, L. Furfaro, and J.M. Dudley, High aspect ratio nanochannel machining using single shot femtosecond Bessel beams, *Appl. Phys. Lett.* **97**, 081102 (2010).

[9] F. Courvoisier, P. A. Lacourt, M. Jacquot, M. K. Bhuyan, L. Furfaro, and J. M. Dudley, Surface nanoprocessing with nondiffracting femtosecond Bessel beams, *Opt. Lett.* **34**, 3163 (2009).

[10] T.H.R. Crawford, A. Borowiec, and H.K. Haugen, Femtosecond laser micromachining of grooves in silicon with 800 nm pulses, *Appl. Phys. A* **80**, 1717 (2005).

[11] M. Toulemonde, S. Unamuno, R. Heddache, M.O. Lampert, M. Hage-Ali, and P. Siffert, Time-resolved reflectivity and melting depth measurements using pulsed ruby laser on silicon, *Appl. Phys. A* **36**, 31 (1985).

[12] T. Grasser, T.-W. Tang, H. Kosina, and S. Selberherr, A Review of Hydrodynamic and Energy-Transport Models for Semiconductor Device Simulation, *Proc. IEEE* **91**, 251 (2003).

[13] J. Bonse, S. M. Wiggins, and J. Solis, Ultrafast phase transitions after femtosecond laser irradiation of indium phosphide, *J. Appl. Phys.* **96**, 2629 (2004).

[14] L.V. Zhigilei, D.S. Ivanov, E. Leveugle, B. Sadigh, and E.M. Bringa, Computer Modeling of Laser Melting and Spallation of Metal Targets, *High Power Laser Ablation V*, *Proc. SPIE* **5448**, 505 (2004).

[15] B. Rethfeld, A. Kaiser, M. Vicanek, G. Simon, Ultrafast Dynamics of Nonequilibrium Electrons in Metals Under Femtosecond Laser Irradiation, *Phys. Rev. B* **65**, 214303 (2002).

[16] N.S. Shcheblanov, T.J. Derrien, T.E. Itina, and C. Phipps, Femtosecond laser interactions with semiconductor and dielectric materials, *AIP Conference Proceedings-American Institute of Physics* **1464**, 1 (2012).

[17] N. Medvedev, B. Rethfeld, Transient dynamics of the electronic subsystem of semiconductors irradiated with an ultrashort vacuum ultraviolet laser pulse, *New J. Phys.* **12**, 073037 (2010).

[18] N. Medvedev, and B. Rethfeld, Effective energy gap of semiconductors under irradiation with an ultrashort VUV laser pulse, *EPL (Europhysics Letters)* **88,** 55001 (2009).

[19] P. Stampfli and K.H. Bennemann, Theory for the instability of the diamond structure of Si, Ge, and C induced by a dense electron-hole plasma, *Phys. Rev.* B **42**, 7163 (1990).

[20] P.L. Silvestrelli, A. Alavi, M. Parrinello, D. Frenkel, *Ab initio* Molecular Dynamics Simulation of Laser Melting of Silicon, *Phys. Rev. Lett.* **77**, 314 (1996).

[21] E.S. Zijlstra, A. Kalitsov, T. Zier, and M.E. Garcia, Fractional diffusion in silicon, *Advanced Materials* **25**, 5605 (2013).

[22] E.S. Zijlstra, A. Kalitsov, T. Zier, and M.E. Garcia, Squeezed thermal phonons precurse nonthermal melting of silicon as a function of fluence, *Physical Review X* **3**, 011005 (2013).

[23] H.O. Jeschke, M.E. Garcia, M. Lenzner, J. Bonse, J. Krüger, and W. Kautek, Laser ablation thresholds of silicon for different pulse durations: theory and experiment, *Appl Surf. Sci.* **197**, 839 (2002).

[24] V.I. Mazhukin, N.M. Nikiforova, and A.A. Samokhin, Photoacoustic Effect upon Material Melting and Evaporation by Laser Pulses, *Phys. Wave Phenom.*, **15**, 81 (2007).

[25] S.I. Anisimov, N. A. Inogamov, A. M. Oparin, B. Rethfeld, T. Yabe, M. Ogawa, and V. E. Fortov, Pulsed laser evaporation: equation-of-state effects, *Appl. Phys. A* **69**, 617 (1999).

[26] M.E. Povarnitsyn, T.E. Itina, M. Sentis, K.V. Khishchenko, and P. R. Levashov, Material decomposition mechanisms in femtosecond laser interactions with metals, *Phys. Rev. B* **75**, 235414 (2007).

[27] S.I. Anisimov, B.L. Kapeliovich, and T.L. Perel'man, Electron emission from metal surfaces exposed to ultrashort laser pulses, *Zh. Eksp. Teor. Fiz.* **66**, 776 (1974) [*Sov. Phys. JETP* **39**, 375 (1974)].

[28] M.E Povarnitsyn, T.E. Itina, K.V. Khishchenko, and P.R. Levashov, Multi-Material Two-Temperature Model for Simulation of Ultra-Short Laser Ablation, *Appl. Surf. Sci.* **253**, 6343 (2007)

[29] G.L. Eesley, Generation of nonequilibrium electron and lattice temperatures in copper by picosecond laser pulses, *Phys. Rev. B* **33**, 2144 (1986).

[30] A.N. Smith, P. M. Norris, Influence of intraband transitions on the electron thermoreflectance response of metals, *Appl. Phys. Lett.* **78**, 1240 (2001).

[31] J.L. Hostetler, A. N. Smith, P. M. Norris, Simultaneous measurement of thermophysical and mechanical properties of thin films, *Int. J. Thermophys.* **19**, 569 (1998).

[32] H.E. Elsayed-Ali, T. Juhasz, G. O. Smith, and W. E. Bron, Femtosecond thermoreflectivity and thermotransmissivity of polycrystalline and single-crystalline gold films, *Phys. Rev. B* **43**, 4488 (1991).

[33] J. L. Hostetler, A. N. Smith, D. M. Czajkowsky, and P. M. Norris, Measurement of the electron-phonon coupling factor dependence on film thickness and grain size in Au, Cr, and Al, *Appl. Optics* **38**, 3614 (1999).

[34] B.A. Remington, G. Bazan, J. Belak, E. Bringa, M. Caturla, J.D. Colvin, M.J. Edwards, S.G. Glendinning, D.S. Ivanov, B. Kad, D.H. Kalantar, M. Kumar, B.F. Lasinski, K.T. Lorenz, J.M. McNaney, D.D. Meyerhofer, M.A. Meyers, S.M. Pollane, D. Rowley, M. Schneider, J.S. Stolken, J.S. Wark, S.V. Weber, W.G. Wolfer, B. Yaakobi, and L.V. Zhigilei, Materials Science Under Extreme Conditions of Pressure and Strain Rate, *Metall. Mater. Trans. A* **35**, 2587 (2004).

[35] Z.H. Jin and K. Lu, Melting of surface-free bulk single crystals, *Phil. Mag. Lett.* **78**, 29 (1998).

[36] M.D. Kluge, J.R. Ray, and A. Rahman, Pulsed laser melting of silicon: A molecular dynamics study, *J. Chem. Phys.* **87**, 2336 (1987).

[37] F.F. Abraham and J.Q. Broughton, Pulsed melting of silicon (111) and (100) surfaces simulated by molecular dynamics, *Phys. Rev. Lett.* **56**, 734 (1986).

[38] H. Häkkinen and U. Landman, Superheating, melting, and annealing of copper surfaces, *Phys. Rev. Lett.* **71**, 1023 (1993).

[39] F.F. Abraham, D.E. Schreiber, M.R. Mruzik, and G.M. Pound, Phase separation in fluid systems by spinodal decomposition: a molecular-dynamics study, *Phys. Rev. Lett*. **36**, 261 (1976).

[40] J.A. Blink and W.G. Hoover, Fragmentation of suddenly heated liquids, *Phys. Rev. A* **32**, 1027 (1985).

[41] Wm. T. Ashurst and B.L. Holian, Droplet formation by rapid expansion of a liquid, *Phys. Rev. E* **59**, 6742 (1999).

[42] L.V. Zhigilei and B.J. Garrison, Pressure waves in microscopic simulations of laser ablation, *Mat. Res. Soc. Symp. Proc*. **538**, 491 (1999).

[43] J.I. Etcheverry and M. Mesaros, Molecular dynamics simulation of the production of acoustic waves by pulsed laser irradiation, *Phys. Rev. B* **60**, 9430 (1999).

[44] R.F.W. Herrmann, J.Gerlach, and E.E.B. Campbell, Ultrashort pulse laser ablation of silicon: an MD simulation study, *Appl. Phys. A* **66**, 35 (1998).

[45] E. Ohmura, I. Fukumoto, and I. Miyamoto, Molecular dynamics simulation of laser ablation of metal and silicon, *Int. J. Japan Soc. Prec. Eng.* **32**, 248 (1998).

[46] L.V. Zhigilei and B.J. Garrison, Microscopic mechanisms of laser ablation of organic solids in the thermal and stress confinement irradiation regimes, *J. Appl. Phys.* **88**, 1281 (2000).

[47] L.V. Zhigilei, Dynamics of the plume formation and parameters of the ejected clusters in short-pulse laser ablation, *Appl. Phys. A* **76**, 339 (2003).

[48] N.A. Inogamov, V.V. Zhakhovsky, Yu.V. Petrov, V.A. Khokhlov, S.I. Ashitkov, K.V. Khishchenko, K.P. Migdal, D.K. Ilnitsky, Yu.N. Emirov, P.S. Komarov, V.V. Shepelev, C.W. Miller, I.I. Oleynik, M.B. Agranat, A.V. Andriyash, S.I. Anisimov and V.E. Fortov, Electron-Ion Relaxation, Phase Transitions, and Surface Nano-Structuring Produced by Ultrashort Laser Pulses in Metals, *Contributions to Plasma Physics* **53**, 796 (2013).

[49] F. Stillinger and T.A. Weber, Computer Simulation of Local Order in Condensed Phases of Silicon, *Phys. Rev. B* **31**, 5262 (1985).

[50] T. Kumagai, S. Izumi, S. Hara, and S. Sakai, Development of bond-order potentials that can reproduce the elastic constants and melting point of silicon for classical molecular dynamics simulation, *Comp. Mat. Sci.* **39**, 457 (2007).

[51] D.S. Ivanov, A.I. Kuznetsov, V.P. Lipp, B. Rethfeld, B.N. Chichkov, M.E. Garcia and W. Schulz, Short laser pulse nanostructuring of metals: direct comparison of molecular dynamics modeling and experiment, *App. Phys. A* **111**, 675 (2013).

[52] B.S. Lee, S. Park, Y.K. Choi, and J.S. Lee, Molecular dynamics study on bulk melting induced by ultrashort pulse laser, *Journal of Mechanical Science and Technology* **25**, 449 (2011).

[53] D.S. Ivanov and L.V. Zhigilei, Combined atomistic-continuum modeling of short-pulse laser melting and disintegration of metal films, *Phys. Rev. B* **68**, 064114 (2003).

[54] D.P. Korfiatis, K.-A. Th. Thoma and J.C. Vardaxoglou, Conditions for femtosecond laser melting of silicon, *J. Phys. D: Appl. Phys.* **40**, 6803 (2007).

[55] H.M. van Driel, Kinetics of high-density plasmas generated in Si by 1.06- and 0.53-μm picosecond laser pulses, *Phys. Rev. B* **35**, 8166 (1987).

[56] T.J. Derrien, T. Sarnet, M. Sentis, and T.E. Itina, Application of a two-temperature model for the investigation of the periodic structure formation on Si surface in femtosecond laser interaction, *Journal of Optoelectronics and Advanced Materials* **12**, 610 (2010).

[57] A. Rämer, O. Osmani, B. Rethfeld, Laser damage in silicon: energy absorption, relaxation and transport, *J. Appl. Phys.* **116**, 053508 (2014).

[58] G.D. Tsibidis, M. Barberoglou, P.A. Loukakos, E. Stratakis, and C. Fotakis, Dynamics of ripple formation on silicon surfaces by ultrashort laser pulses in subablation conditions, *Phys. Rev. B* **86**, 115316 (2012).

[59] Y. Gan and J.K. Chen, A hybrid method for integrated atomistic-continuum simulation of ultrashort-pulse laser interaction with semiconductors, *Computer Physics Communications* **183**, 278 (2012).

[60] P. Lorazo, L.J. Lewis, M. Meunier, Thermodynamic pathways to melting, ablation, and solidification in absorbing solids under pulsed laser irradiation, *Phys. Rev. B* **73**, 134108 (2006).

[61] L. Shokeen, P.K. Schelling, Role of electronic-excitation effects in the melting and ablation of laser-excited silicon, *Comp. Mat. Sci.* **67,** 316 (2013).

[62] V.P. Lipp, D.S. Ivanov, B. Rethfeld, M.E. Garcia, On the interatomic interaction potential that describes bond weakening in classical molecular-dynamic modeling, *Journal of Optical Technology* **81**, 254 (2014).

[63] J.F. Young, H.M. Van Driel, Ambipolar diffusion of high-density electrons and holes in Ge, Si, and GaAs: Many-body effects, *Phys. Rev. B* **26**, 2147 (1982).

[64] A.B. Sproul and M.A. Green, Improved value for the silicon intrinsic carrier concentration from 275 to 375 K, *J. Appl. Phys.* **70**, 846 (1991).

[65] J. Crank, P. Nicolson, A practical method for numerical evaluation of solutions of partial differential equations of the heat conduction type, *Mathematical Proceedings of the Cambridge Philosophical Society* **43**, 50 (1947).

[66] Tuncer Cebeci, Convective Heat Transfer, Springer, ISBN 0-9668461-4-1 (2002).

[67] V.P. Lipp, D.S. Ivanov, B. Rethfeld, M.E. Garcia, Semi-implicit finite-difference method with predictor-corrector for solution of diffusion equation with non-linear source terms, *in preparation for Comp. Phys. J.*

[68] A.D. Bristow, N. Rotenberg, and H. M. Van Driel, Two-photon absorption and Kerr coefficients of silicon for 850–2200 nm, *Appl. Phys. Lett.* **90**, 191104 (2007).

[69] A.A. Ionin, S.I. Kudryashov, L.V. Seleznev, D.V. Sinitsyn, A.F. Bunkin, V.N. Lednev, S.M. Pershin, Thermal melting and ablation of silicon by femtosecond laser radiation, *Zh. Eksp. Teor. Fiz.* **143**, 403 (2013) [*J. Exp. Theor. Phys.* **116**, 347 (2013)].

[70] J. Bonse, All-optical characterization of single femtosecond laser-pulse-induced amorphization in silicon, *Appl. Phys. A: Mater. Sci. Proc. A* **84**, 63 (2006).

[71] J. Bonse, S. Baudach, J. Krüger, W. Kautek, M. Lenzner, Femtosecond laser ablation of silicon–modification thresholds and morphology, *Appl. Phys. A* **74**, 19 (2002).

[72] J. Bonse, K.-W. Brzezinka and A.J. Meixner, Modifying single-crystalline silicon by femtosecond laser pulses: an analysis by micro Raman spectroscopy, scanning laser microscopy and atomic force microscopy, *Appl. Surf. Sci.* **221**, 215 (2004).

[73] E. J. Yoffa, Dynamics of dense laser-induced plasmas, *Phys. Rev. B* **21**, 2415 (1980).

[74] G.E. Jellison and F. A. Modine, Optical functions of silicon between 1.7 and 4.7 eV at elevated temperatures, *Phys. Rev. B* **27**, 7466 (1983).

[75] J.K. Chen, D.Y. Tzou, and J.E. Beraun, Numerical Investigation of Ultrashort Laser Damage in Semiconductors, *Int. J. Heat and Mass Transfer* **48**, 501 (2005).

[76] M.P. Allen and D.J. Tildesley, Computer Simulation of Liquids, Oxford University Press, New York, 1987.

[77] J. Tersoff, Empirical interatomic potential for silicon with improved elastic properties, *Phys. Rev. B* **38**, 9902 (1988).

[78] S.J. Cook and P. Clancy, Comparison of semi-empirical potential functions for silicon and germanium, *Phys. Rev. B* **47**, 7686 (1993).

[79] M. I. Baskes, Modified embedded-atom potentials for cubic materials and impurities, *Phys. Rev. B* **46**, 2727 (1992).

[80] S. Ryu and W. Cai, Comparison of thermal properties predicted by interatomic potential models, *Modelling Simul. Mater. Sci. Eng.* **16**, 085005 (2008).

[81] V.S. Dozhdikov, A.Yu. Basharin, and P.R. Levashov, Two-phase simulation of the crystalline silicon melting line at pressures from −1 to 3 GPa, *J. Chem. Phys.* **137**, 054502 (2012).

[82] M.H. Grabow, G.H. Gilmer and A.F. Bakker, Molecular Dynamics Studies of Silicon Solidification and Melting, *MRS Proceedings* **141**, 349 (1988).

[83] S.G. Volz and G. Chen, Molecular-dynamics simulation of thermal conductivity of silicon crystals, *Phys. Rev. B* **61**, 2651 (2000).

[84] C.A. da Cruz, K. Termentzidis, P. Chantrenne, and X. Kleber, Molecular dynamics simulations for the prediction of thermal conductivity of bulk silicon and silicon nanowires: Influence of interatomic potentials and boundary conditions, *J. Appl. Phys.* **110**, 034309 (2011).

[85] H. Balamane, T. Halicioglu, and W.A. Tiller, Comparative study of silicon empirical interatomic potentials, *Phys. Rev. B* **46**, 2250 (1992).

[86] J. Q. Broughton, X. P. Li, Phase diagram of silicon by molecular dynamics, *Phys. Rev. B* **35**, 9120 (1987).

[87] H. Balamane, T. Halicioglu, and W. A. Tiller, Comparative study of silicon empirical interatomic potentials, *Phys. Rev.* **B** 46, 2250 (1992).

[88] A. Rousse, C. Rischel, S. Fourmaux, I. Uschmann, S. Sebban, G. Grillon, Ph. Balcou, E. Förster, J.P. Geindre, P. Audebert, J.C. Gauthier, and D. Hulin, Non-thermal melting in semiconductors measured at femtosecond resolution, *Nature* **410**, 65 (2001).

[89] M. Head-Gordon, J.C. Tully, Molecular dynamics with electronic frictions, *J. Chem. Phys.* **103**, 10137 (1995).

[90] D.M. Duffy and A.M. Ruthersford, Including the effects of electronic stopping and electron–ion interactions in radiation damage simulations, *J. Phys.: Condens. Matter* **19**, 016207 (2007).

[91] E. Zarkadoula, S. Daraszewicz, D.M. Duffy, M. Seaton, I.T. Todorov, K. Nordlund, M.T. Dove, and K. Trachenko, Electronic effects in high-energy radiation damage in iron, *J. Phys.: Condens. Matter* **26**, 085401 (2014).

[92] L.V. Zhigilei and D.S. Ivanov, Channels of energy redistribution in short-pulse laser interactions with metal targets, *Appl. Surf. Sci.* **248**, 433 (2005).

[93] D. Errandonea, The melting curve of ten metals up to 12 GPa and 1600 K, *J. Appl. Phys.* **108**, 033517 (2010).

[94] B. Rethfeld, K. Sokolowski-Tinten, D. von der Linde, and S.I. Anisimov, Ultrafast thermal melting of laser-excited solids by homogeneous nucleation, *Phys. Rev. B* **65**, 092103 (2002).

[95] L.V. Zhigilei, Z. Lin, and D.S. Ivanov, Atomistic Modeling of Short Pulse Laser Ablation of Metals: Connections between Melting, Spallation, and Phase Explosion, *J. Phys. Chem. C* **113**, 11892 (2009).

[96] W.F. Gale, and T.C. Totemeier, eds. Smithells metals reference book. Butterworth-Heinemann, 2003.

[97] D.S. Ivanov and L.V. Zhigilei, Effect of Pressure Relaxation on the Mechanisms of Short-Pulse Laser Melting, *Phys. Rev. Lett.* **91**, 105701 (2003).

[98] E. Leveugle, D.S. Ivanov, and L.V. Zhigilei, Photomechanical spallation of molecular and metal targets: molecular dynamics study, *Appl. Phys. A* **79**, 1643 (2004).

[99] S. Ryu and W. Cai, Comparison of thermal properties predicted by interatomic potential models, *Modelling and Simulation in Materials Science and Engineering* **16**, 085005 (2008).

[100] M.H. Grabow, G.H. Gilmer, and A.F. Bakker, Molecular dynamics studies of silicon solidification and melting, *Mat. Res. Soc. Symp. Proc.* **141**, 349 (1989)

[101] Weast, R. C. (Ed.), Handbook of Chemistry and Physics, 90th ed. CRC.

[102] R.F. Wood and G. E. Giles, Macroscopic theory of pulsed-laser annealing. I. Thermal transport and melting, *Phys. Rev. B* **23**, 2923 (1981).

[103] D. Agassi, Phenomenological model for picosecond pulse laser annealing of semiconductors, *J. Appl. Phys.* **55**, 4376 (1984).

[104] C.D. Thurmond, The Standard Thermodynamic Functions for the Formation of Electrons and Holes in Ge, Si, GaAs, and GaP, *J. Electrochem. Soc.* **122**, 1133 (1975).

[105] R. Vankemmel, W. Schoenmaker, and K. De Meyer, A unified wide temperature range model for the energy gap, the effective carrier mass, and intrinsic concentration in silicon, *Solid State Electron.* **36**, 1379 (1993).

[106] G. E. Jellison and F. A. Modine, Optical absorption of silicon between 1.6 and 4.7 eV at elevated temperatures, *Appl. Phys. Lett.* **41**, 180 (1982).

[107] J. Dwiezor and W. Schmid, Auger coefficients for highly doped and highly excited silicon, *Appl. Phys. Lett.* **31**, 346 (1977).

[108] J. Geist and W.K. Gladden, Transition rate for impact ionization in the approximation of a parabolic band structure, *Phys. Rev. B* **27**, 4833 (1983).

[109] J.R. Meyer, M.R. Kruer, and F.J. Bartoli, Optical heating in semiconductors: Laser damage in Ge, Si, InSb, and GaAs, *J. Appl. Phys.* **51**, 5513 (1980).

[110] Ioffe Physical Technical Institute, Electronic Archive "New Semiconductor Materials. Characteristics and Properties", http://www.ioffe.rssi.ru/SVA/NSM/Semicond/Si/index.html.

[111] E. Yu. Tonkov, Phase Transformations of Elements under High Pressure (Metallurgiya, Moscow, 1988; CRC Press, Boca Raton, Florida, United States, 2005).

[112] V.M. Glazov and O.D. Shchelikov, Volume Changes during Melting and Heating of Silicon and Germanium Melts, *High Temp.* **38**, 405 (2000).

[113] Qi Zhang, Qikai Li, and Mo Li, Melting and superheating in solids with volume shrinkage at melting: A molecular dynamics study of silicon, *J. Chem. Phys.* **138**, 044504 (2013).

[114] R.A. Logan and W.L. Bond, Density Change in Silicon upon Melting, *J. Appl. Phys.* **30**, 3 (1959).

[115] H. Watanabe, N. Yamada, and M. Okaji, Linear thermal expansion coefficient of silicon from 293 to 1000 K, *International journal of thermophysics* **25**, 221 (2004).

[116] A.S. Okhotin, A.S. Pushkarskii, and V.V. Gorbachev, Thermophysical Properties of Semiconductors, Moscow, "Atom" Publ. House, 1972 (in Russian); http://www.ioffe.ru/SVA/NSM/Semicond/Si/thermal.html.

[117] C.C. Yang, J.C. Li, and Q. Jiang, Temperature–pressure phase diagram of silicon determined by Clapeyron equation, *Solid state communications* **129**, 437 (2004).

[118] J.R. Morris, X. Song, The melting lines of model systems calculated from coexistence simulations, *J. Chem. Phys.* **116**, 9352 (2002).

[119] P.J. Steinhardt, D.R. Nelson, and M. Ronchetti, Bond-orientational order in liquids and glasses, *Phys. Rev. B* **28**, 784 (1983).